\documentclass[journal]{IEEEtran}
\usepackage{graphicx}
\usepackage{arydshln}
\usepackage{multirow}
\usepackage{booktabs}
\usepackage{setspace}
\usepackage{amsmath}
\usepackage{amssymb}
\usepackage{algorithm}
\usepackage{algorithmicx}
\usepackage{cite}
\usepackage{epstopdf}
\usepackage{color}
\usepackage{cases}
\usepackage[caption=false]{subfig}

\usepackage{mathtools}
\usepackage{scalerel,stackengine}
\usepackage[noend]{algpseudocode}

\stackMath

\newsavebox{\foobox}

\newcommand{\vect}[1]{\boldsymbol{\mathbf{#1}}}

\newcommand{\EQ}{\begin{equation}\begin{array}{lllllllll}}
\newcommand{\EE}{\end{array}\end{equation}}
\newcommand{\MT}{\left[ \begin{array}{ccccccccc}}
	\newcommand{\EM}{\end{array}\right]}

\newcommand{\m}{\boldsymbol}
\newcommand{\bmat}[1]{\begin{bmatrix} #1 \end{bmatrix}}

\def\BState{\State\hskip-\ALG@thistlm}
\newcommand{\normof}[1]{\|#1\|}

\expandafter\def\expandafter\normalsize\expandafter{%
	\normalsize
	\setlength\abovedisplayskip{3pt}
	\setlength\belowdisplayskip{3pt}
	\setlength\abovedisplayshortskip{3pt}
	\setlength\belowdisplayshortskip{3pt}
}

\begin{document}
	
	\title{
		Comparing Kalman Filters and Observers for Power System Dynamic State Estimation with Model Uncertainty and Malicious Cyber Attacks 
	}
	
	\author{Junjian~Qi,~\IEEEmembership{Senior Member,~IEEE,}
	        Ahmad F. Taha,~\IEEEmembership{Member,~IEEE,}
	        and Jianhui Wang,~\IEEEmembership{Senior Member,~IEEE}
	        \thanks{
	        J.~Qi is with the Department of Electrical and Computer Engineering, University of Central Florida, Orlando, FL 32816 USA (e-mail: Junjian.Qi@ucf.edu).
	        
	        A. F. Taha is with the Department of Electrical and Computer Engineering, the University of Texas at San Antonio, San Antonio, TX 78249 
	        (e-mail: ahmad.taha@utsa.edu).
	        
	        J. Wang is with the Department of Electrical Engineering at Southern Methodist University, Dallas, TX 75275 USA 
	        (e-mail: jianhui.wang@ieee.org).
	        }
	}

	\maketitle

	\begin{abstract}
		Kalman filters and observers are two main classes of dynamic state estimation (DSE) routines. Power system DSE has been implemented by various Kalman filters, such as the extended Kalman filter (EKF) and the unscented Kalman filter (UKF). In this paper, we discuss two challenges for an effective power system DSE: (a) model uncertainty and (b) potential cyber attacks. To address this, the cubature Kalman filter (CKF) and a nonlinear observer are introduced and implemented. Various Kalman filters and the observer are then tested on the 16-machine, 68-bus system given realistic scenarios under model uncertainty and different types of cyber attacks against synchrophasor measurements. It is shown that CKF and the observer are more robust to model uncertainty and cyber attacks than their counterparts. Based on the tests, a thorough qualitative comparison is also performed for Kalman filter routines and observers. 
	\end{abstract}


	\section{Introduction}
	
	\IEEEPARstart{S}{tate} estimation is a crucial application in the energy management system (EMS). 
	The well-known static state estimation (SSE) methods \cite{se3,se2,se5,qi2012power} assume that the power system is operating in quasi-steady state, based on which the static states---the voltage magnitude and phase angles of the buses---are estimated by using SCADA and/or synchrophasor measurements. SSE is critical for power system monitoring as it provides inputs for other EMS applications such as automatic generation control and optimal power flow. 
	
	However, SSE may not be sufficient for desirable situational awareness as the system states evolve more rapidly due to an increasing penetration of renewable generation and distributed energy resources. Therefore, dynamic state estimation (DSE) processes estimating the dynamic states (i.e., the internal states of generators) by using highly synchronized PMU measurements with high sampling rates will be critical for the wide-area monitoring, protection, and control of power systems. 
	
	For both SSE and DSE, two significant challenges make their practical application significantly difficult. 
	First, the system model and parameters used for estimation can be inaccurate, which is often called \textit{model uncertainty} \cite{zhou1996robust}, consequently deteriorating estimation in some scenarios. Second, the measurements used for estimation are vulnerable to cyber attacks, which in turn leads to compromised measurements that can greatly mislead the estimation. 
	
	For the first challenge, there are recent efforts on validating the dynamic model of the generator and calibrating its parameters 
	\cite{huang2013generator,ariff2015estimating}, which DSE can be based on. 
	However, model validation itself can be very challenging. Hence, it is a more viable solution to improve the estimators by making them more robust to the model uncertainty.
	
	For the second challenge, false data injection attacks against SSE are proposed in~\cite{liu2011false}.  
	After that it has been widely studied 
	about how to mitigate this type of attack and further secure the monitoring and control of power grids 
	\cite{6032057,6194239,6490324}. 
	
	As for the approaches for performing DSE, there are mainly two classes of methods that have been proposed:
	
	\begin{enumerate}
		\item \textit{Stochastic Estimators}: given a discrete-time representation of a dynamical system, the observed measurements, and the statistical information on process noise and measurement noise, Kalman filter (KF) and its many derivatives have been proposed that calculate the Kalman gain as a function of the relative certainty of the  current state estimate and the measurements \cite{kalman1960new,jazwinski2007stochastic,ukf,ukf1,CKF}.
		\item \textit{Deterministic Observers}: given a continuous- or discrete-time dynamical system depicted by state-space matrices, a combination of matrix equalities and inequalities are solved, while guaranteeing asymptotic (or bounded) estimation error. The solution to these equations is often matrices that are used in an observer to estimate states and other dynamic quantities \cite{kang2013survey,Radke2006,Hidayat2011}.
	\end{enumerate}
	
	For power systems, DSE has been implemented by several stochastic estimators, such as extended Kalman filter (EKF) 
	\cite{ekf1,ekf2}, unscented Kalman filter (UKF) \cite{valverde2011unscented,ghahremani2011online,pwukf1,pwukf2,sun2016power}, 
	square-root unscented Kalman filter (SR-UKF)~\cite{van2001square,qi,qi2015dynamic}, 
	extended particle filter \cite{zhou,cui2015particle}, and ensemble Kalman filter \cite{zhou2015dynamic}. 
	While these techniques produce good estimation under nominal conditions, most of them lack the ability to deal with significant model uncertainty and malicious cyber attacks. 

	In order to improve the robustness of KFs, a generalized maximum-likelihood-type estimate is proposed in \cite{gandhi2010robust} and a two-stage KF is proposed in \cite{zhang2014two}. 
	Besides, iterated EKF \cite{zhao1}, $H_\infty$ EKF \cite{zhao2}, and robust UKF \cite{zhao3} are have also been developed for power system DSE. 

The goal of this paper is to present alternatives that address these 
limitations. The contributions are summarized as follows. 
	
	First, we design a nonlinear observer for the power system DSE problem that only requires computing a Luenberger-like gain matrix. This computation can be performed offline---and hence the presented observer is scalable for large-scale power networks. The designed observer requires obtaining scalar parameters that depict or bound the nonlinearities arising from the power system model. Numerical algorithms are provided to find these scalar parameters, in comparison with the observer design literature that obtains these scalars analytically which is impractical for large-scale power networks with high nonlinearities. The observer is endowed with the the following properties and virtues: (a) It considers a nonlinear measurement model, in comparison with other methods that utilize linearized measurement models; (b)  assumes that the generators' control inputs are not known to the state estimation method; (c) tolerates three classes of cyber-attacks (data integrity, denial of service, and replay attack) and other disturbances while accurately reconstructing the power system state within seconds of an attack or large disturbance; (d) assumes no statistical properties of the noise targeting process and measurement models; (e) requires no major real-time computation, in comparison with other estimation methods that are computationally expensive. To our knowledge, this contribution is the first of its kind in the power system DSE literature 
	in comparison with Kalman filter derivatives.

Second, we introduce cubature Kalman filter (CKF) \cite{CKF} that uses a more accurate cubature approach and possesses an important virtue of mathematical rigor rooted in the third-degree spherical-radial cubature rule for numerically computing Gaussian-weighted integrals. Without a stem at the center in the cubature-point set, CKF does not have the numerical instability problem of UKF \cite{CKF,qi2015dynamic}.  
	
{Last but not least, we design a realistic power system DSE problem by developing the system and measurement models and considering various practical scenarios such as unknown initial conditions, model uncertainties including process noise, unknown and unavailable inputs, and inaccurate parameters, and different types of measurement noises and cyber attacks against measurements. We present thorough numerical experiments to showcase the performance of the nonlinear observer and CKF in comparison with three other methods that have been recently applied to DSE. 
The conceptual strengths and limitations of different methods with significant model uncertainty and cyber attacks are also discussed.
	
	The remainder of this paper is organized as follows. 
	The physical 
	depictions of the model uncertainty and attack-threat model are introduced in Section~\ref{sec:at}. The CKF and the nonlinear observer are introduced in Sections~\ref{sec:CKF1} and \ref{sec:Observers}.
	Then, numerical results are given in Section~\ref{sec:NumericalResults}. Finally, insightful remarks and conclusions are presented in Sections~\ref{sec:remarks} and~\ref{sec:conc}.

	\allowdisplaybreaks
	\section{Nonlinear Multi-Machine Power System Model} ~\label{sec:multimachinedynamics}
	Here we briefly discuss the power system model used for DSE. 
	Each of the $G$ generators is described by the fourth-order transient model in local $\mathrm{d}-\mathrm{q}$ reference frame:
	\begin{eqnarray}~\label{gen model}
	\hspace{-0.3cm}\left\{
	\arraycolsep=1.4pt\def\arraystretch{1.8}
	\begin{array}{ll}
	\dot{\delta_i}=\omega_i-\omega_0 \\
	\dot{\omega}_i=\frac{\omega_0}{2H_i}\Big(T_{\mathrm{m}i}-T_{\mathrm{e}i}-\frac{K_{\mathrm{D}i}}{\omega_0}(\omega_i-\omega_0)\Big) \\
	\dot{e}'_{\mathrm{q}i}=\frac{1}{T'_{\mathrm{d0}i}}\Big(E_{\mathrm{fd}i}-e'_{\mathrm{q}i}-(x_{\mathrm{d}i}-x'_{\mathrm{d}i})i_{\mathrm{d}i}\Big) \\
	\dot{e}'_{\mathrm{d}i}=\frac{1}{T'_{\mathrm{q0}i}}\Big(-e'_{\mathrm{d}i}+(x_{\mathrm{q}i}-x'_{\mathrm{q}i})i_{\mathrm{q}i}\Big),
	\end{array}
	\right.
	\end{eqnarray}
	where $i$ is the generator serial number, $\delta_i$ is the rotor angle,
	$\omega_i$ is the rotor speed in rad/s, and $e'_{\mathrm{q}i}$ and $e'_{\mathrm{d}i}$ are the transient voltage along $\mathrm{q}$ and $\mathrm{d}$ axes; 
	$i_{\mathrm{q}i}$ and $i_{\mathrm{d}i}$ are stator currents at $\mathrm{q}$ and $\mathrm{d}$ axes;
	$T_{\mathrm{m}i}$ is the mechanical torque, $T_{\mathrm{e}i}$ is the electric air-gap torque, 
	and $E_{\mathrm{fd}i}$ is the internal field voltage; $\omega_0$ is the rated value of angular frequency, $H_i$ is the inertia constant, and $K_{\mathrm{D}i}$ is the damping factor; 
	$T'_{\mathrm{q0}i}$ and $T'_{\mathrm{d0}i}$ are the open-circuit time constants for $\mathrm{q}$ and $\mathrm{d}$ axes; $x_{\mathrm{q}i}$ and $x_{\mathrm{d}i}$ are the synchronous reactance and $x'_{\mathrm{q}i}$ and $x'_{\mathrm{d}i}$ are the transient reactance respectively at the $\mathrm{q}$ and $\mathrm{d}$ axes. 
	
	The $T_{\mathrm{m}i}$ and $E_{\mathrm{fd}i}$ in (\ref{gen model}) are considered as inputs. 
	The set of generators where PMUs are installed is denoted by $\mathcal{G}_{\mathrm{P}}$. 
	For generator $i \in \mathcal{G}_{\mathrm{P}}$,  the terminal voltage phaosr $E_{\mathrm{t}i}=e_{\mathrm{R}i}+je_{\mathrm{I}i}$ and current phasor $I_{\mathrm{t}i}=i_{\mathrm{R}i}+ji_{\mathrm{I}i}$ can be measured and are used as the outputs. Correspondingly, the state vector $\m{x} \in \mathbb{R}^n$, input vector $\m{u} \in \mathbb{R}^v$, and output vector $\m{y} \in \mathbb{R}^p$ are
	\begin{subequations}
	\begin{align}
	\m{x} &= \big[\m{\delta}^\top \quad \m{\omega}^\top \quad \m{e'_{\textbf{q}}}^\top \quad \m{e'_{\textbf{d}}}^\top\big]^\top \\
	\m{u} &= \big[\m{T_\textbf{m}}^\top \quad \m{E_{\textbf{fd}}}^\top\big]^\top \\
	\m{y} &= \big[\m{e_\textbf{R}}^\top \quad \m{e_\textbf{I}}^\top \quad \m{i_\textbf{R}}^\top \quad \m{i_\textbf{I}}^\top\big]^\top.
	\end{align}
	\end{subequations} 
	
	The $T_{\mathrm{e}i}$, $i_{\mathrm{d}i}$, and $i_{\mathrm{q}i}$ can be written as functions of $\m{x}$:
	\begin{subequations} \label{temp}
	\begin{align}
	\Psi_{\mathrm{R}i}&=e'_{\mathrm{d}i}\sin\delta_i+e'_{\mathrm{q}i}\cos\delta_i \\
	\Psi_{\mathrm{I}i}&=e'_{\mathrm{q}i}\sin\delta_i-e'_{\mathrm{d}i}\cos\delta_i \\
	I_{\mathrm{t}i}&=\m{\overline{Y}}_i(\m{\Psi}_{\textbf{R}}+j\m{\Psi}_{\textbf{I}}) \\
	i_{\mathrm{R}i}&= \operatorname{Re}(I_{\mathrm{t}i}) \label{iR} \\
	i_{\mathrm{I}i}&= \operatorname{Im}(I_{\mathrm{t}i}) \label{iI} \\
	i_{\mathrm{q}i}&=\frac{S_\mathrm{B}}{S_{\mathrm{N}i}}(i_{\mathrm{I}i}\sin\delta_i+i_{\mathrm{R}i}\cos\delta_i) \\
	i_{\mathrm{d}i}&=\frac{S_\mathrm{B}}{S_{\mathrm{N}i}}(i_{\mathrm{R}i}\sin\delta_i-i_{\mathrm{I}i}\cos\delta_i) \\
	e_{\mathrm{q}i}&=e'_{\mathrm{q}i}-x'_{\mathrm{d}i}i_{\mathrm{d}i} \\
	e_{\mathrm{d}i}&=e'_{\mathrm{d}i}+x'_{\mathrm{q}i}i_{\mathrm{q}i} \\
	T_{\mathrm{e}i}&=\frac{S_\mathrm{B}}{S_{\mathrm{N}i}}(e_{\mathrm{q}i}i_{\mathrm{q}i}+e_{\mathrm{d}i}i_{\mathrm{d}i}),
	\end{align}
	\end{subequations}
	where $\Psi_i=\Psi_{\mathrm{R}i}+j\Psi_{\mathrm{I}i}$ is the voltage source, $\m{\Psi}_\textbf{R}$ and $\m{\Psi}_\textbf{I}$ are column vectors of all generators' $\Psi_{\mathrm{R}i}$ and $\Psi_{\mathrm{I}i}$, $e_{\mathrm{q}i}$ and $e_{\mathrm{d}i}$ are the terminal voltage at $\mathrm{q}$ and $\mathrm{d}$ axes, $\m{\overline{Y}}_i$ is the $i$th row of the admittance matrix of the reduced network $\m{\overline{Y}}$, and $S_\mathrm{B}$ and $S_{\mathrm{N}i}$ are the system base MVA and the base MVA for generator $i$, respectively. 
	
	In \eqref{temp}, the outputs $i_{\mathrm{R}i}$ and $i_{\mathrm{I}i}$ have been written as functions of $\m{x}$.
	Similarly, the outputs $e_{\mathrm{R}i}$ and $e_{\mathrm{I}i}$ can also be written as function of $\m{x}$:
	\begin{subequations} \label{eReI}
	\begin{align}
	e_{\mathrm{R}i}&= e_{\mathrm{d}i}\sin\delta_i+e_{\mathrm{q}i}\cos\delta_i \label{eR} \\
	e_{Ii}&=e_{\mathrm{q}i}\sin\delta_i-e_{\mathrm{d}i}\cos\delta_i. \label{eI}
	\end{align}
	\end{subequations} 
	
	The dynamic model (\ref{gen model}) can then be rewritten in a general state space form as
	\begin{eqnarray} \label{model1}
	\hspace{-0.3cm}\left\{
	\arraycolsep=1.4pt\def\arraystretch{1.1}
	\begin{array}{ll}
	\dot{\m{x}} &= \m{A}\m{x}+\m{B}\m{u}+\m{\phi}(\m{x}) \\
	\m{y} &= \m{h}(\m{x}),
	\end{array}
	\right.
	\end{eqnarray}
	where
	\begingroup
	\small
	\begin{align}
	\m A = \left[
	\renewcommand{\arraystretch}{2}
	\begin{array}{c:c:c:c}
	\,\,\,\,\,\,\,\,\,\,\,\,\,& \m{I}_G & & \\ \hdashline
	 & \big(-{\m{K}_\mathbf{D}}{\oslash}{2{\m H}}\big)^\mathrm{d} & & \\ \hdashline
	 & & \big(-{\m 1_G}\oslash{\m{T}'_{\mathbf{d0}}}\big)^\mathrm{d} & \\ \hdashline
	 & & & \big(-{\m 1_G}\oslash {\m{T}'_{\mathbf{q0}}}\big)^\mathrm{d}
	\end{array}\right], \notag
	\end{align}
	\begin{equation}
	\m{B}=\left[
	\renewcommand{\arraystretch}{2}
	\begin{array}{c:c}
	 &  \\ \hdashline
	\big({\omega_0 \m 1_G}\oslash {2{\m H}}\big)^\mathrm{d} & \\ \hdashline
	 & \big({\m 1_G}\oslash {\m{T}'_{\mathbf{d0}}}\big)^\mathrm{d} \\ \hdashline
	 & \\
	\end{array}\right], \; \notag
	\end{equation}
	\begin{equation}
	\m{\phi}=\left[
	\renewcommand{\arraystretch}{1.3}
	\begin{array}{c}
	-\omega_0 \m{1}_G \\
	\big({\omega_0 \m 1_G} \oslash {2{\boldsymbol H}}\big)\circledast \big(-\m{T}_\mathbf{e} + \m{K}_\mathbf{D} \m{1}_G \big) \\ 
	\big({\m 1_G} \oslash {\m{T}'_{\mathbf{d0}}}\big)\circledast \big( -(\m{x}_\mathbf{d} - \m{x}'_\mathbf{d}) \m{i}_{\mathbf{d}} \big) \\
	\big({\m 1_G} \oslash {\m{T}'_{\mathbf{q0}}}\big) \circledast \big( (\m{x}_\mathbf{q} - \m{x}'_\mathbf{q}) \m{i}_{\mathbf{q}} \big)
	\end{array}\right], \notag
	\end{equation} \endgroup
	and $\m{h}$ include functions (\ref{iR})--(\ref{iI}) and (\ref{eReI}) for all generators, $\oslash$ and $\circledast$ are the Hadamard division/product (elementwise division/product) of a vector, 
	and $(\m a)^\mathrm{d}$ gets a square diagonal matrix with the elements of vector $\m a$ on the main diagonal. 
	
	Note that the model presented here is used for DSE for which the real-time inputs are assumed to be unavailable 
	and $T_{\mathrm{m}i}$ and $E_{\mathrm{fd}i}$ only take steady-state values, mainly because these inputs are difficult to measure \cite{ekf2,pwukf2}. 
	However, when we simulate the power system to mimic the real system dynamics, we model an IEEE Type DC1 excitation system and a simplified turbine-governor system
	for each generator and thus $T_{\mathrm{m}i}$ and $E_{\mathrm{fd}i}$ change with time due to the governor and the excitation control, which leads to a tenth order generator model. 
	More details about the model can be found in \cite{7491374}. 
	
	We do not directly use a detailed model including the exciter and governor as in \cite{cui2015particle} for the DSE 
	mainly because 1) A good model should be simple enough to facilitate design \cite{zhou1996robust}, 
	2) it is harder to validate a detailed model and there are also more parameters that need to be calibrated \cite{huang2013generator,hajnoroozi2015generating,ariff2015estimating}, 
	and 3) the computational burden can be higher for a more detailed model, which may not satisfy the requirement of real-time estimation.

	\section{Model Uncertainty and Cyber Attacks}~\label{sec:at}
	
	The dynamic model of the power system can be written in a general state space form as
	\begin{subnumcases}{\label{model_ori}}
	\vect{x}=\vect{f}(\vect{x},\vect{u}) \label{model-1} \\
	\vect{y}=\vect{h}(\vect{x},\vect{u}), \label{model-2}
	\end{subnumcases}
	where $\vect{x} \in \mathbb{R}^n$, $\vect{u} \in \mathbb{R}^v$, and $\vect{y} \in \mathbb{R}^p$ are the vectors of the state, input, and output, and $\vect{f}$ and $\vect{h}$ are the nonlinear state transition functions and measurement functions.
	We rewrite (\ref{model_ori}) by separating the nonlinear term in the state transition functions as
	\begin{subnumcases}{\label{model}}
	\dot{\vect{x}} = \vect{A}\vect{x}+\vect{B}\vect{u}+\vect{\phi}(\vect{x}) \\
	\vect{y} = \vect{h}(\vect{x},\vect{u}),
	\end{subnumcases}
	where $\vect{\phi}(\vect{x})$ represents the nonlinear term that models the interconnections in a multi-machine power system. 
	
Two great challenges for an effective DSE are the model uncertainty and potential cyber attacks---discussed next.
	\subsection{Model Uncertainty}
	The term \textit{model uncertainty} refers to the differences or errors between models and reality. 
	Various control and estimation theory studies investigated methods that addresses the discrepancy between the actual physics and models.
	The model uncertainty can be caused by the following reasons.
	
	\begin{enumerate}
		\item \textit{Unknown inputs}:
		The unknown inputs against the system dynamics include $\vect u_\mathrm{d}$~(representing the unknown plant disturbances), $\vect u_\mathrm{u}$ (denoting the unknown control inputs), and $\vect{f}_\mathrm{a}$~(depicting potential generators actuator faults). For simplicity, we can combine them into one unknown input quantity 
		$\vect w = \bmat{\vect u_\mathrm{d}^{\top} & \vect u_\mathrm{u}^{\top} & \vect f_\mathrm{a}^{\top}}^{\top}$. 
		Defining $\vect{B}_\mathrm{w}$ to be the known weight distribution matrix of the distribution of unknown inputs with respect to each state-equation. 
		The term $\vect{B}_\mathrm{w}\vect w$ models a general class of unknown inputs such as: nonlinearities, modeling uncertainties, noise, parameter variations, unmeasurable system inputs, model reduction errors, and actuator faults~\cite{Chen2012,Pertew2005}. 
		The process dynamics under unknown inputs can be written as follows:
		\begin{equation}~\label{equ:plantDynamics}
		\dot{\vect x} = \vect{A} \vect x + \vect{B}  \vect u + \vect{B}_\mathrm{w}  \vect w  + \vect{\phi}(\vect x).
		\end{equation}
		\item \textit{Unavailable inputs}: Real-time inputs $\vect u$ 
		can be unavailable, in which case the steady-states inputs $\vect u_0$ are used for estimation. 
		\item \textit{Parameter inaccuracy}: The parameters in the system model can be inaccurate. For example, the reduced admittance matrix 
		can be inaccurate when a fault or the following topology change are not detected. 
	\end{enumerate}

	\subsection{Cyber Attacks} \label{model cyber}
	The National Electric Sector Cybersecurity Organization Resource (NESCOR) developed cyber-security failure scenarios with corresponding impact analyses~\cite{NESCOR2014}.	
	The WAMPAC failure scenarios motivate the research in this paper include: a) \textit{Measurement Data (from PMUs) Compromised due to PDC Authentication Compromise} and b) \textit{Communications Compromised between PMUs and Control Center}~\cite{NESCOR2014}. 
	Specifically, we consider the following three types of attacks \cite{NESCOR2014,sridhar2012cyber}. 
	
	\begin{enumerate}
		\item \textit{Data integrity attacks}: An adversary attempts to corrupt the content of either the measurement or the control signals. 
		A specific example of data integrity attacks are Man-in-the-Middle attacks, where the adversary intercepts the measurement signals and modifies them in transit.
		For DSE the PMU measurements can be modified and corrupted. 
		\item \textit{Denial of Service (DoS) attack}: An attacker attempts to introduce a denial in communication of measurement. 
		The communication of a sensor could be jammed by flooding the network with spurious packets. 
		For DSE the consequence can be that the updated measurements cannot be sent to the control center.
		\item \textit{Replay attacks}: A special case of data integrity attacks, where the attacker replays a previous snapshot of a valid communication packet sequence that contains measurements in order to deceive the system. 
		For DSE the PMU measurements can be changed to be those in the past.
	\end{enumerate}

	
	Apart from the above-mentioned cyber attacks against the PMU measurement, the commonly assumed Gaussian distribution of the PMU measurement noise may not hold for real data. Extensive results using field PMU data from WECC system has revealed that the Gaussian assumption is questionable \cite{error}. Therefore, it would be valuable to evaluate the performance of different DSE methods under non-Gaussian noise.

	\section{Kalman Filters for Power System DSE}~\label{sec:CKF1}

	
	Unlike many estimation methods that are either computationally unmanageable or require special assumptions about the form of the process and
	observation models, KF only utilizes the first two moments of the state (mean and covariance) in its update rule \cite{kalman1960new}. 
	It consists of two steps: in prediction step, the filter propagates the estimate from last time step to current time step; 
	in update step, the filter updates the estimate using collected measurements. 
	KF was initially developed for linear systems while for power system DSE the system equations and outputs have strong nonlinearity. 
	Thus variants of KF that can deal with nonlinear systems have been introduced, such as EKF and UKF. 
	

	\subsection{Extended Kalman Filter}
	
	Although EKF maintains the elegant and computationally efficient recursive update form of KF, it works well only in a `mild' nonlinear environment,  
	owing it to the first-order Taylor series approximation for nonlinear functions \cite{CKF}. 
	It is sub-optimal and can easily lead to divergence.
	Also, the linearization can be applied only if the Jacobian matrix exists and calculating Jacobian matrices can be difficult and error-prone.
	For DSE, EKF has been discussed in \cite{ekf1,ekf2}.

	\subsection{Unscented Kalman Filter} \label{UKF}
	
	The unscented transformation (UT) \cite{ut} is developed to address the deficiencies of linearization 
	by providing a more direct and explicit mechanism for transforming mean and covariance information. 
	Based on UT, Julier et al. \cite{ukf,ukf1} propose the UKF as a derivative-free alternative to EKF. 
	The Gaussian distribution is represented by a set of deterministically chosen sample points called sigma points. 
	The UKF has been applied to power system DSE, for which no linearization or calculation of Jacobian matrices is needed \cite{valverde2011unscented,ghahremani2011online,pwukf1,pwukf2,sun2016power}.
	
	In UKF, a total of $2\,n+1$ sigma points (denoted by $\boldsymbol{\mathcal{X}}$) are calculated from the columns of the matrix $\eta\sqrt{\boldsymbol{P}}$ as 
	\begin{subnumcases}{\label{sigma}}
	\boldsymbol{\mathcal{X}}^{(0)}=\boldsymbol{m} \\
	\boldsymbol{\mathcal{X}}^{(i)}=\boldsymbol{m}+\left[\eta\sqrt{\boldsymbol{P}}\right]_i,\quad i=1,\ldots,n   \\
	\boldsymbol{\mathcal{X}}^{(i)}=\boldsymbol{m}-\left[\eta\sqrt{\boldsymbol{P}}\right]_i,\quad i=n+1,\ldots,2n 
	\end{subnumcases}
	with weights
	\begin{subnumcases}{\label{weight}}
	\boldsymbol{w}_m^{(0)}=\frac{\lambda}{n+\lambda}  \\
	\boldsymbol{w}_c^{(0)}=\frac{\lambda}{n+\lambda}+(1-\alpha^2+\beta)  \\
	\boldsymbol{w}_m^{(i)}=\frac{1}{2(n+\lambda)}, \quad i=1,\ldots,2n  \\
	\boldsymbol{w}_c^{(i)}=\frac{1}{2(n+\lambda)}, \quad i=1,\ldots,2n,
	\end{subnumcases}
	where the matrix square root of a positive semidefinite matrix $\boldsymbol{P}$ is a matrix $\boldsymbol{S}=\sqrt{\boldsymbol{P}}$ such that $\boldsymbol{P}=\boldsymbol{S}\boldsymbol{S}^\top$, $\boldsymbol{w}_m$ and $\boldsymbol{w}_c$ are respectively weights for the mean and the covariance, $\eta=\sqrt{n+\lambda}$, $\lambda$ is a scaling parameter defined as $\lambda=\alpha^2(n+\kappa)-n$, and $\alpha$, $\beta$, and $\kappa$ are constants and $\alpha$ and $\beta$ are nonnegative.
	
	The basic idea of UKF is to choose the sigma-point set to capture a number of low-order moments of the prior density of the states 
	as correctly as possible, and then compute the 
	posterior statistics of the nonlinear functions 
	(either state transition functions $\vect{f}$ or measurement functions $\vect{h}$) by UT which approximates the mean and the covariance of the nonlinear function by a weighted sum of projected sigma points. 

	However, for the sigma-points, the stem at the center (the mean) is highly significant as it carries more weight which is usually negative for high-dimensional systems. 
	Therefore, the UKF is supposed to encounter numerical instability troubles when used in high-dimensional problems. 
	Several techniques including the square-root unscented Kalman filter (SR-UKF) have been proposed to solve this problem \cite{van2001square,qi}. Recently SR-UKF has been applied to DSE in power systems in \cite{qi2015dynamic}.

	\subsection{Cubature Kalman Filter}~\label{sec:CKFF}
	EKF and UKF can suffer from the curse of dimensionality while becoming detrimental in high-dimensional state-space models of size twenty or more---especially when there are high degree of nonlinearities in the equations that describe the state-space model \cite{CKF,bellman}, which is exactly the case for power systems. 
	Making use of the spherical-radial cubature rule, Arasaratnam et al. \cite{CKF} propose CKF, 
	which possesses an important virtue of mathematical rigor rooted in the third-degree spherical-radial cubature rule for numerically computing Gaussian-weighted integrals. 
	In this paper we will apply CKF to power system DSE. 
	Compared with EKF, UKF, and SR-UKF, CKF has the following advantages:
	\begin{enumerate}
		\item Compared with EKF and similar to UKF and SR-UKF, CKF is also derivative-free and is easier for application. 
		\item Similar to UKF and SR-UKF, CKF also uses a weighted set of symmetric points to approximate the Gaussian distribution. 
		But the cubature-point set 
		does not have a stem at the center 
		and thus does not have the numerical instability problem of UKF discussed in Section \ref{UKF}. 
		\item 
		UKF treats the derivation of the sigma-point set for the prior density and the computation fo posterior statistics as two disjoint problems. 
		By contrast, 
		CKF directly derives the cubature-point set to accurately compute the first two-order moments of a nonlinear transformation, 
		therefore naturally increasing the accuracy of the numerical estimates for moment integrals \cite{CKF}. 
		\item As suboptimal Bayesian filters, EKF, UKF, and CKF all have some robustness to model uncertainties and measurement outliers \cite{gadsden2014combined}. 
		The extent of robustness depends on their ability to accurately deal with the nonlinear transformations. 
		The EKF is the least robust method due to a first-order Taylor series approximation of the nonlinear functions while the CKF has the highest robustness thanks to its more accurate cubature approach, 
		which will be validated in the result section. 
	\end{enumerate}

	\section{Nonlinear Observers for Power System DSE}~\label{sec:Observers}
	
	Dynamic observers have been thoroughly investigated for different classes of systems. To mention a few, they have been developed for linear time-invariant (LTI) systems, nonlinear time-invariant (NLTI) systems, LTI and NLTI systems with unknown inputs, sensor and actuator faults, stochastic dynamical systems, and 
	hybrid systems~\cite{kang2013survey,Radke2006}. 
	
	Most observers utilize the plant's outputs and inputs to generate real-time estimates of the plant states, unknown inputs, and sensor faults. 
	The cornerstone is the innovation function---sometimes a simple gain matrix designed to nullify the effect of unknown inputs and faults. Linear and nonlinear functional observers, sliding-mode observers, unknown input observers, and observers for fault detection and isolation are all examples on developed observers for different classes of systems, under different assumptions~\cite{Hidayat2011}.
	
	In comparison with KF techniques, nonlinear and robust observers have not been utilized for power system DSE. 
	However, they inherently possess the theoretical, technical, and computational capabilities to perform good estimation of the power system's dynamic states.
	As for implementation, observers are simpler than KFs. For observers, matrix gains are computed offline to guarantee the asymptotic stability of the estimation error or the boundedness of the estimation error within a neighborhood of the origin.
	
	Here, we present a recently developed observer in~\cite{Zhang2012} that can be applied for DSE in power systems. 
	This observer assumes that the nonlinear function $\vect\phi(\vect x)$ in (\ref{model}) satisfies the one-sided Lipschitz condition. 
	Specifically, there exists $\rho \in \mathbb{R}$ such that $\forall\, \vect x_1,\vect x_2$ in a region $D$ including the origin with respect to the state $\vect x$, there is
	$$\langle \vect\phi(\vect x_1)-\vect\phi(\vect x_2),\vect x_1 - \vect x_2\rangle \leq \rho \, \normof{\vect x_1 - \vect x_2}^2, $$	
	where $\langle \cdot,\cdot \rangle$ is the inner product. Besides, the nonlinear function is also assumed to be quadratically inner-bounded as
	$$\big( \vect\phi(\vect x_1)-\vect\phi(\vect x_2)\big)^{\top} \big( \vect\phi(\vect x_1)-\vect\phi(\vect x_2)\big) \leq \mu\, \normof{\vect x_1 -\vect x_2}^{2} $$ $$ + \varphi \,\langle  \vect\phi(\vect x_1)-\vect\phi(\vect x_2),\vect x_1 - \vect x_2\rangle, $$
	where $\mu$ and $\varphi$ are real numbers. Similar results related to the dynamics of multi-machine power systems established a similar quadratic bound on the nonlinear component (see~\cite{Siljak2002}). 
	To determine the constants $\rho, \mu, \,\textrm{and}\, \varphi$, a simple offline algorithm can be implemented. For example, 
	we can define a \textit{region of interest} $\mathcal{D} \subset \mathbb{R}^{n}$ to be the state-space region where the system operates.
	For the multi-machine power network, this region is the intersection of all upper and lower bounds of states, which can be written as
	$$\mathcal{D} = [\vect x_1^{\min}, \vect x_1^{\max}] \times [\vect x_2^{\min}, \vect x_2^{\max}]\times \cdots \times [\vect x_n^{\min}, \vect x_n^{\max}].$$  
	This $\mathcal{D}$ can be obtained by the method discussed in \cite{qi2017nonlinear}. 
	We sample random points in this region. 
	Denser sampling yields a more realistic Lipschitz constant, while requiring more computational time. 
	Let $n_{\mathcal{D}}$ be the total number of samples inside $\mathcal{D}$. Algorithm~\ref{algoLC} includes the steps required to obtain $\rho$. 
	Specifically, $\rho$ can be calculated from
	\begin{align*}
	\rho = \lim \sup \Big( \beta \Big(  \frac{\partial \vect \phi}{\partial \vect x}  \Big) \Big)
	\end{align*}
	for all $\vect x \in \mathcal{D}$, where $\beta(\vect H)$ denotes the logarithmic matrix norm of matrix $\vect H$ defined as
	\begin{align*}
	\beta(\vect H) = \displaystyle \lim_{\epsilon\to0} \frac{\Vert \vect I+ \epsilon \vect H \Vert -1}{\epsilon},
	\end{align*}
	where $\Vert \cdot\Vert$ represents any matrix norm. It is shown in~\cite{vidyasagar2002nonlinear} that the logarithmic matrix norm can also be written as
	$$ \beta(\vect H) = \lambda_{\max}\left( \frac{1}{2}\left(\vect H + \vect H^{\top}\right) \right) \leq \normof{\vect H}.$$
	At each iteration, we obtain the maximum eigenvalue of 
	$ \dfrac{1}{2}\left(\dfrac{\partial \vect\phi(\vect x)}{\partial \vect x}  +\left( \dfrac{\partial \vect\phi(\vect x)}{\partial \vect x}\right)^{\top}\right) $
	where the Jacobian of the nonlinear function is evaluated at 
	the $i$th sampled point. Finally, $\rho$ is computed by finding the maximum value of $\beta(\cdot)$ over $\mathcal{D}$. 
	\begin{algorithm}[!t]
		\caption{Obtaining One-Sided Lipschitz Constant $\rho$}\label{algoLC}
		\begin{algorithmic}
			\State \textbf{input} $\m\phi(\vect x)$ and $\mathcal{D}$
			\State $\rho_0 \leftarrow  -\infty $ 
			\For{$i=1:n_{D}$}
			\State $\vect x \leftarrow \vect x_i$ 
			\State \textbf{compute} 
			$\rho_i =   \left[ \lambda_{\max} \left(\dfrac{1}{2}\left(\dfrac{\partial \vect \phi(\vect x)}{\partial \vect x}  +\left(\dfrac{\partial \m\phi(\vect x)}{\partial \vect x}\right)^{\top}\right)\right) \right] $
			\State $\rho_i = \max(\rho_{i-1}, \rho_i)$
			\EndFor
			\State \textbf{end for}
			\State \textbf{output} $\rho \leftarrow \rho_{n_{\mathcal{D}}}$
		\end{algorithmic}
	\end{algorithm}
	
	To compute the quadratic inner-boundedness constants $\mu$ and $\varphi$, a similar algorithm can be obtained. In particular, instead of sampling over individual $\vect x_i \in \mathcal{D}$, two state-space samples $\vect x_i$ and $\vect x_j$ can be sampled at each iteration $(i,j)$, and 
	$$\big( \vect\phi(\vect x_i)-\vect\phi(\vect x_j)\big)^{\top} \big( \vect\phi(\vect x_i)-\vect\phi(\vect x_j)\big) \leq \mu_{i,j}\, \normof{\vect x_i -\vect x_j}^{2} $$ $$ + \varphi_{i,j} \,\langle  \vect\phi(\vect x_i)-\vect\phi(\vect x_j),\vect x_i - \vect x_j\rangle, $$
	is evaluated iteratively for all possible permutations $\vect x_i$ and $\vect x_j$ in $\mathcal{D}$ to obtain the maximum values for $\mu$ and $\varphi$ that satisfy the above inequality.

	
	Following these assumptions, the dynamics of this observer can be written as
	\begin{equation}~\label{equ:ObserverTwo}
	\stackon[.5pt]{\vect{\hat{x}}}{.}= \vect A \stackon[.5pt]{\vect{\hat{x}}}{} +\vect B \vect u+\vect\phi(\stackon[.5pt]{\vect{\hat{x}}}{})+\vect L \big(\vect y-\vect C \stackon[.5pt]{\vect{\hat{x}}}{} \big),
	\end{equation}
	where $\vect L$ is a matrix gain determined by Algorithm~\ref{algoOBS2}.
	
	\begin{algorithm}[!h]
		\small	\caption{Observer Design Algorithm}\label{algoOBS2}
		\begin{algorithmic} 
			\State \textbf{compute} constants $\rho, \mu,$ and $\varphi$ via an offline search algorithm
			\State \textbf{solve} this LMI for $\epsilon_1, \epsilon_2, \sigma > 0$ and $\vect P=\vect P^{\top} \succ \textit{\textbf{O}}$:
			\begin{equation}~\label{equ:LMIObserverDes}
			\hspace*{-0.4cm} \left[
			\renewcommand{\arraystretch}{2.1}
			\begin{array}{c;{2pt/2pt}c}
			\vect A^{\top}\vect P+\vect P \vect A+(\epsilon_1\rho+\epsilon_2\mu)\vect I_n \\-\sigma\,\vect C^{\top}\vect C & \vect P+\dfrac{\varphi\,\epsilon_2-\epsilon_1}{2}\vect I_n \\\hdashline[2pt/2pt] 
			\left(\vect P+\dfrac{\varphi\,\epsilon_2-\epsilon_1}{2}\vect I_n\right)^{\top} & -\epsilon_2 \vect I_n
			\end{array}
			\right]
			< 0.
			\end{equation}
			\State \textbf{obtain} the observer design gain matrix $\vect L$:
			\begin{equation} \label{L}
			\vect L=\frac{\sigma}{2}\vect P^{-1}\vect C^{\top}.
			\end{equation}
			\State \textbf{simulate} the observer design given in~\eqref{equ:ObserverTwo}
		\end{algorithmic}
	\end{algorithm}
	First, given the Lipschitz constants $\rho,\,\varphi$, and $\mu$, the linear matrix inequality in~\eqref{equ:LMIObserverDes} is solved for positive constants $\epsilon_1, \epsilon_2$, and $\sigma$ and a symmetric positive semi-definite matrix $\vect P$. Utilizing the $\vect L$ in~\eqref{L}, the state estimates generated from~\eqref{equ:ObserverTwo} are guaranteed to converge to the actual values of the states.
	
	Note that the observer design utilizes linearized measurement functions $\vect C$, which for power system DSE can be obtained by linearizing the nonlinear functions in (\ref{model}). 
	However, since the measurement functions have high nonlinearity, when performing the estimation we do not use (\ref{equ:ObserverTwo}), as in~\cite{Zhang2012}, 
	but choose to directly use the nonlinear measurement functions as 
	\begin{equation}~\label{equ:obsdyn}\vect L
	\stackon[.5pt]{\vect{\hat{x}}}{.} = \vect A \stackon[.5pt]{\vect{\hat{x}}}{} +\vect B\vect u+\vect\phi(\stackon[.5pt]{\vect{\hat{x}}}{})+\vect L \big(\vect y-\vect h(\stackon[.5pt]{\vect{\hat{x}}}{})\big).
	\end{equation}
	The observer presented here is designed to seamlessly deal with model uncertainties. 
	It assumes that the nonlinearities present in the power system dynamics (i.e., $\boldsymbol{\phi}(\vect{x})$) satisfy the quadratic inner-boundedness and the one-sided Lipschitz condition. 
	As we illustrate in the results, power systems in fact satisfy these conditions (see~\cite{Siljak2002}). 
	Then any predicted uncertainty in the model, in addition to unknown inputs, can be added to the nonlinearity function $\boldsymbol{\phi}(\vect{x})$. In addition to this contribution, we emphasize that the numerical algorithms given to find the constants $\rho, \varphi,$ and $\mu$ can be used to any nonlinear dynamic system since analytically computing these constants is impractical for large-scale networks with a huge number of nonlinearities.
	
	The main idea behind the observer is to minimize the difference between the estimated measurements (i.e., $\stackon[.5pt]{\vect{\hat{y}}}{}(t)$) and the actual ones ($\vect{y}(t)$) through the innovation term $\vect{L}\big(\vect{y}-\vect{h}(\stackon[.5pt]{\vect{\hat{x}}}{})\big)$. The objective of this term is to nullify/minimize the discrepancies due to errors in the estimation, model uncertainties, measurement noise, or attack vectors. The difference between $\vect{y}(t)$ and $\stackon[.5pt]{\vect{\hat{y}}}{}(t)$ yields an estimate for the attack vector. 
	Hence, the states evolution for the observer are indirectly aware of the differences between measured and potentially corrupt outputs and the estimated ones. 
	Given the solution to the LMIs, the estimation error dynamics will be asymptotically stable, which implies that even under attack vectors, the observer will succeed in providing converging state estimates. Finally, it is important to mention that Algorithm ~\ref{algoOBS2} can be performed offline, which implies that the observer in real-time only requires a state-estimate update while all other quantities are given; after finding $\vect L$ one can simulate \eqref{equ:obsdyn} without needing to perform other computations. 

	\section{Numerical Results}~\label{sec:NumericalResults}
	
	Here we test EKF, UKF, SR-UKF, CKF, and the nonlinear observer on the 16-machine 68-bus system extracted from Power System Toolbox (PST) \cite{chow1992toolbox}.
	For the DSE we consider both unknown inputs to the system dynamics and cyber attacks against the measurements including data integrity, DoS, and 
	replay attacks; see Section~\ref{sec:at}. All tests are performed on a 3.2-GHz Intel(R) Core(TM) i7-4790S desktop.
	
	For simulating the power system to mimic the real system dynamics, we model an IEEE Type DC1 excitation system and a simplified turbine-governor system, 
	which leads to a $10$th order generator model. More details about the model can be found in \cite{7491374}. 
	The simulation data is generated as follows.
	\begin{enumerate}
		\item The simulation data is generated by the detailed 10-th order model. The sampling rate is $60$ samples/s.
		\item In order to generate dynamic response, a three-phase fault is applied at bus $6$ of branch $6-11$ 
		and is cleared at the near and remote ends after $0.05$ and $0.1$ s. 
		\item All generators are equipped with PMUs at their terminal buses. The real and imaginary parts of the voltage phasor and current phasor are considered as measurements. 
		\item The sampling rate of the measurements is set to be $60$ frames/s to mimic the PMU sampling rate. 
		\item Gaussian process noise is added and the 
		process noice covariance is a diagonal matrix, whose diagonal entries are the square of 5\% of the largest state changes \cite{zhou}.
		\item Gaussian noise with variance $0.01^2$ is added to the PMU measurements. 
		\item Each entry of the unknown input coefficients $\vect{B}_\mathrm{w}$ is a random number that follows normal distribution with zero mean and variance as the square of 50\% of the largest state changes.
		Note that the variance here is much bigger than that of the process noise.
		\item The unknown input vector $\vect w$ is set as a function of $t$ as
		$$ \vect w(t) =\bmat{
			0.5\cos(\omega_ut)\\
			0.5\sin(\omega_ut)\\
			0.5\cos(\omega_ut)\\
			0.5\sin(\omega_ut)\\
			-e^{-5t}\\
			0.2\,e^{-t}\cos(\omega_u t)\\
			0.2\cos(\omega_ut)\\
			0.1\sin(\omega_ut)
		},$$
		where $\omega_u=100$ is the frequency of the given signals. The unknown inputs are manually chosen, showing different scenarios for inaccurate model and parameters without a 
		predetermined 
		distribution. 
	\end{enumerate}
	
	For DSE we use the fourth-order generator model in \cite{qi2015dynamic,qi}. We do not use a more detailed model 
	mainly because 1) A good model should be simple enough to facilitate design \cite{zhou1996robust}, 
	2) it is harder to validate a more detailed model and there are also more parameters that need to be calibrated \cite{huang2013generator,ariff2015estimating}, and 3) the computational burden can be higher for a more detailed model, which may not satisfy the requirement of real-time estimation. 
	The Kalman filters and the observer are set as follows. 
	\begin{enumerate}
		\item DSE is performed on the post-contingency system on time period $[0,10\,\textrm{s}]$, which starts from the fault clearing.
		\item The initial estimated mean of the rotor speed is set to be $\omega_0$ and that for the other states is set to be twice of the real initial states. 
		\item The initial estimation error covariance is set to be $0.1\vect{I}_n$.
		\item As mentioned before, the covariance of the process noise is set as a diagonal matrix, whose diagonal entries are the square of 5\% of the largest state changes \cite{zhou}.
		\item The covariance for the measurement noise is a diagonal matrix, whose diagonal entries are $0.01^2$, as in \cite{zhou}. 
		\item For both UKF and SR-UKF, $2n+1$ sigma points are used in the unscented transformation. 
		\item For UKF and SR-UKF, a popular heuristic $n+\kappa=3$ proposed in \cite{julier2000new} is used to choose the parameter $\kappa$ in unscented transformation in order to minimize the moments of the standard Gaussian and the sigma points up to the fourth order. 
		\item For UKF is performed by using the EKF/UKF toolbox \cite{hartikainen2011optimal}, in which the function `$\operatorname{schol}$' is used to calculate the lower triangular Cholesky factor of a matrix and can get an output even when the matrix is not positive semidefinite \cite{qi2015dynamic}. 
		\item For the observer in Section~\ref{sec:Observers}, the LMI~\eqref{equ:LMIObserverDes} is solved via CVX on MATLAB~\cite{CVX1}. 
		The Lipschitz constants in Algorithm~\ref{algoOBS2} are set as $\rho=10, \mu=1,$ and $\varphi=1$. 
		\item The mechanical torque and internal field voltage are considered as unavailable inputs and take steady-state values, because they are difficult to measure \cite{ekf2,pwukf2}.  
		\item On $[0,1\,\textrm{s}]$ the reduced admittance matrix 
		is the one for the pre-contingency state.
		\item Data integrity, DoS, and replay attacks, as discussed in Section \ref{model cyber}, are added to the PMU measurements. 
	\end{enumerate}
	

	\subsection{Scenario 1: Data Integrity Attack}
	
	Data integrity attack is added to the first eight measurements, i.e., the real parts of the voltage phasors. 
	The compromised measurements are obtained by scaling the real measurements by $0.6$ and $1/0.6$, respectively, for the first four and the last four measurements.
	The 2-norm of the relative error of the states, $||(\vect x(t)-\hat{\vect x}(t))/\vect x(t)||_2$, for different estimation methods is shown in Fig. \ref{sce1_norm}. 
	It is seen that the error norm for both CKF and the observer can quickly converge among which the observer converges faster, while the value that CKF converges to is slightly smaller in magnitude. 
	By contrast, EKF, UKF, and SR-UKF do not perform as well. 
	
	\begin{figure}[!t]
		\centering
		\includegraphics[width=2.4in]{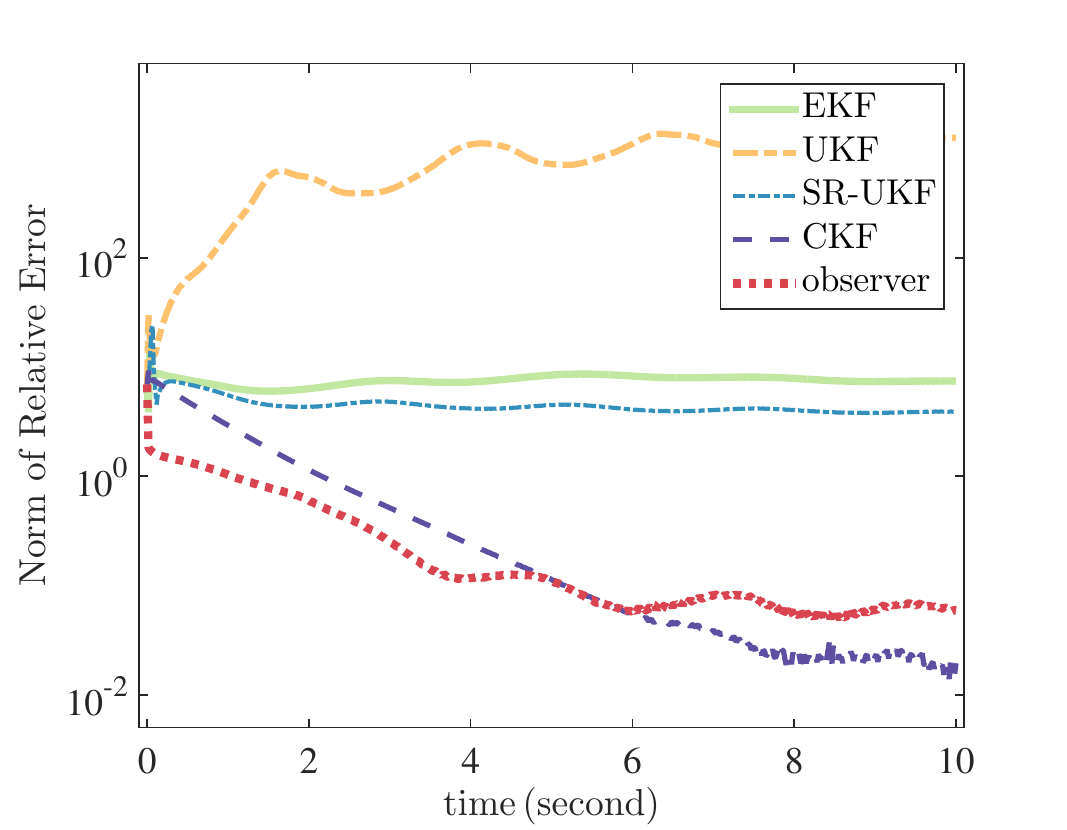}
		\caption{Norm of relative error of the states in Scenario 1.}
		\label{sce1_norm}
	\end{figure}

	We also show the states estimation for Generator 1 in Fig. \ref{sce1_states}. 
	It is seen that the observer and CKF converge rapidly while the EKF fails to converge after $10$ seconds.
	The estimation for UKF is separately shown in Fig. \ref{sce1_states_ukf} because its estimated states are far away from the real states. 
	Note that the real system dynamics are stable while the UKF estimation misled by the data integrity attack indicates that the system is unstable.

	\begin{figure}[!t]
		\centering
		\includegraphics[width=3.3in]{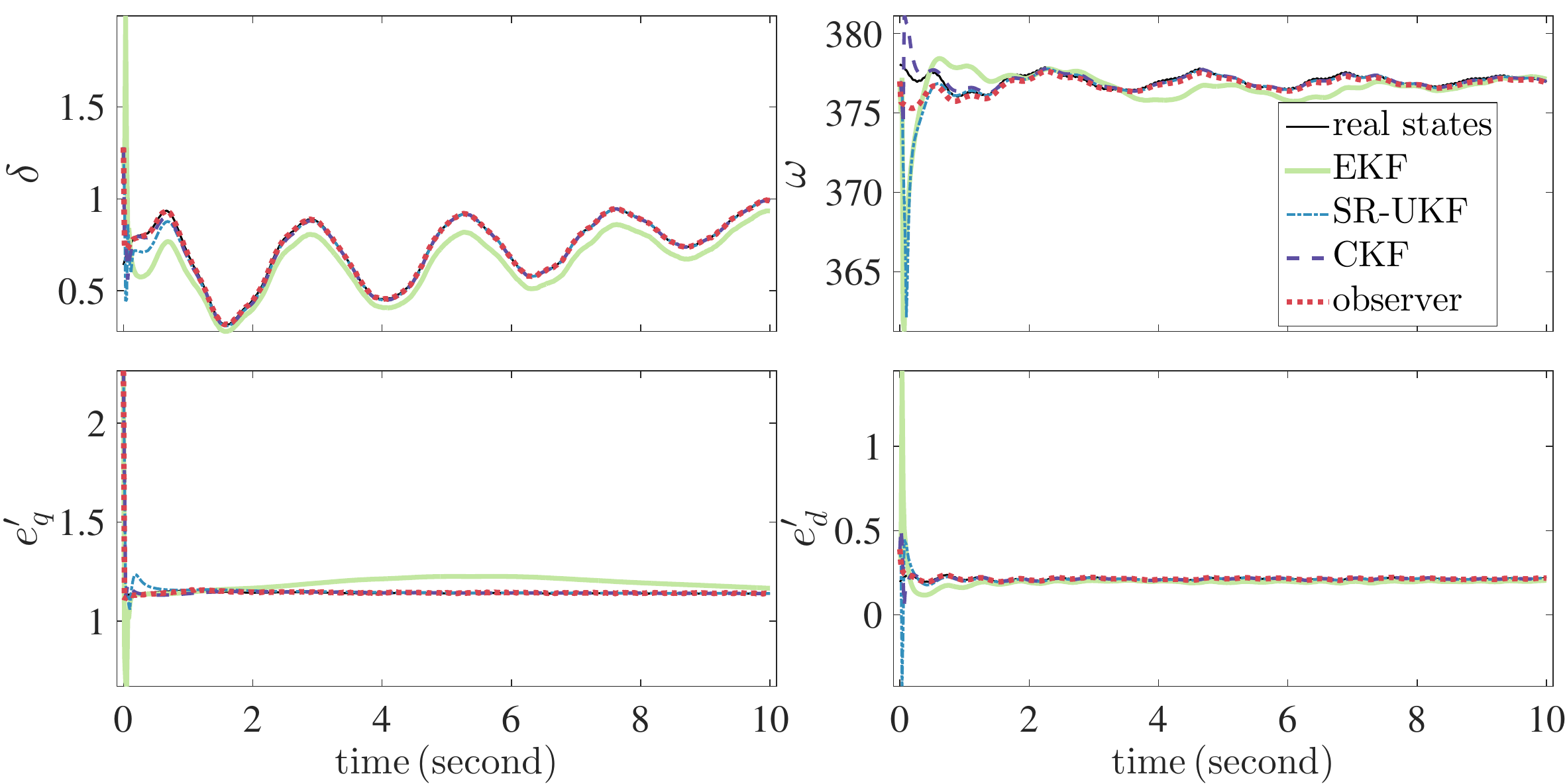}
		\caption{Estimated states by EKF, SR-UKF, CKF, and observer in Scenario 1.}
		\label{sce1_states}
	\end{figure}
	
	\begin{figure}[!t]
		\centering
		\includegraphics[width=3.3in]{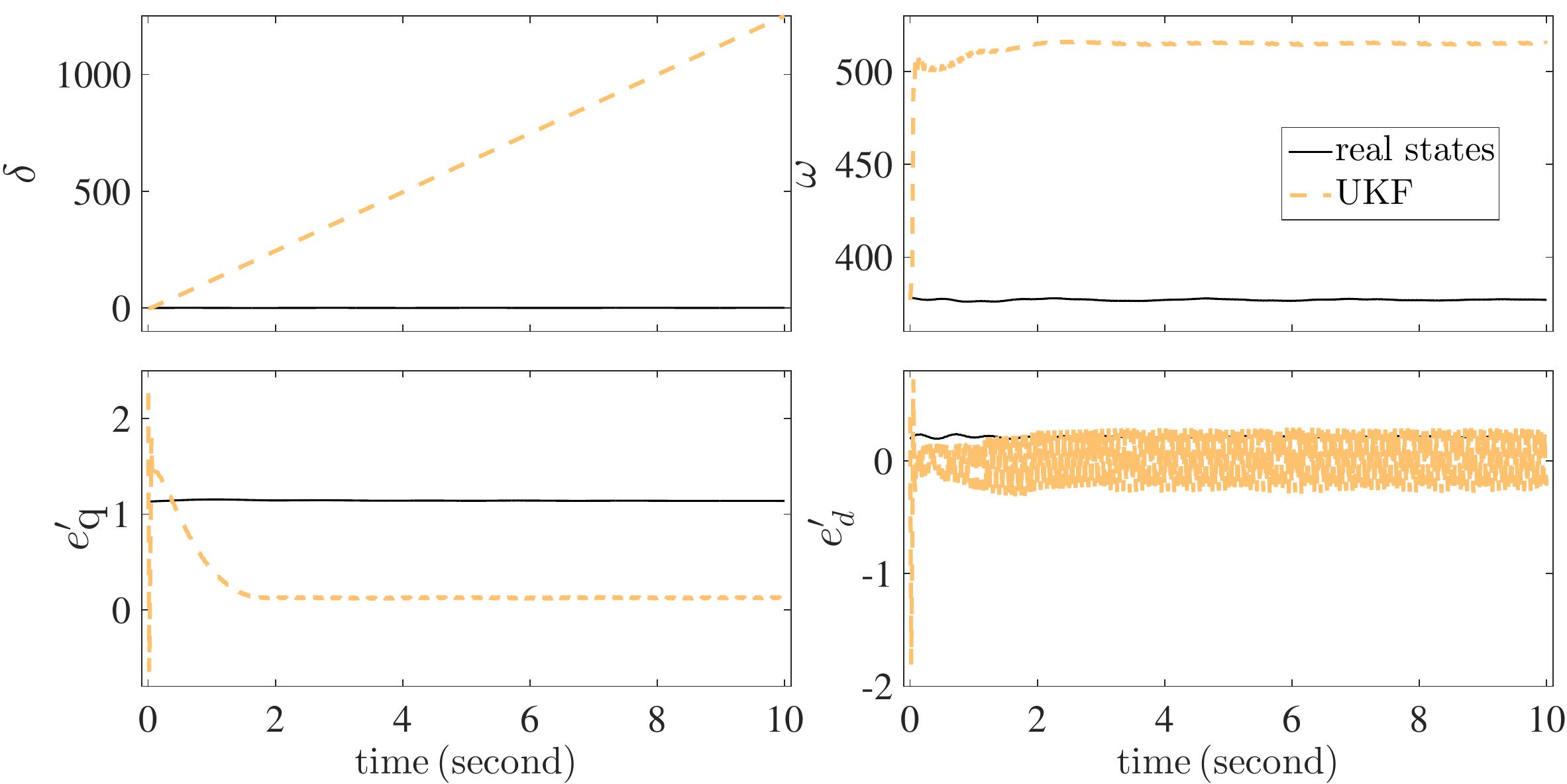}
		\caption{Estimated states by UKF in Scenario 1.}
		\label{sce1_states_ukf}
	\end{figure}
	
	The real, compromised, and estimated values for the first measurement 
	are shown in Fig.~\ref{sce1_meas}. 
	For the observer, CKF, and SR-UKF, the estimated measurements are very close to the actual ones. 
	For EKF there are some differences between the estimates and the real values, while 
	UKF's generated estimates are close to the compromised measurements, indicating that it is completely misled by the cyber attack.

	\begin{figure}[!t]
		\centering
		\includegraphics[width=2.4in]{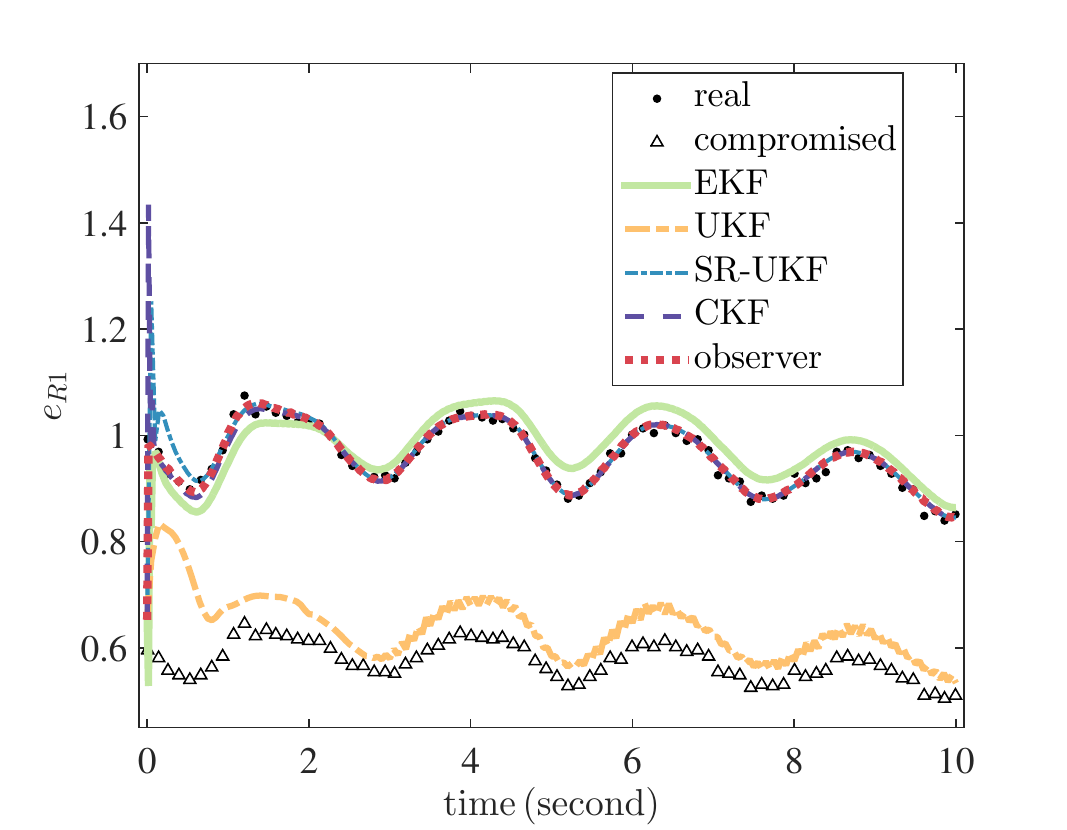}
		\caption{Estimated values for the first measurement in Scenario 1.}
		\label{sce1_meas}
	\end{figure}

	\subsection{Scenario 2: DoS Attack and Scenario 3: Replay Attack}
	
	The first eight measurements are kept unchanged for $t \in [3\,\textrm{s},\,6\,\textrm{s}]$ to mimic the DoS attack in which case the updated measurements cannot be sent to the control center due to, for example, jammed communication between PMU to PDC or between PDC to the control center
	\cite{NESCOR2014}. 
	
	Replay attack is added on the first eight measurements for which there is $y_i(t)=y_i(t-3)$ for $t\in [3\,\textrm{s},\,6\,\textrm{s}]$. 
	
	The 2-norm of the relative error of the states is shown in Fig. \ref{sce2_3_norm} and the results are very similar to those in Scenario 1.
	
	
	\begin{figure}[!t]
		\begin{minipage}{.5\linewidth}
			\centering
			\subfloat[]{\label{sce2_norm}\includegraphics[width=1.81in]{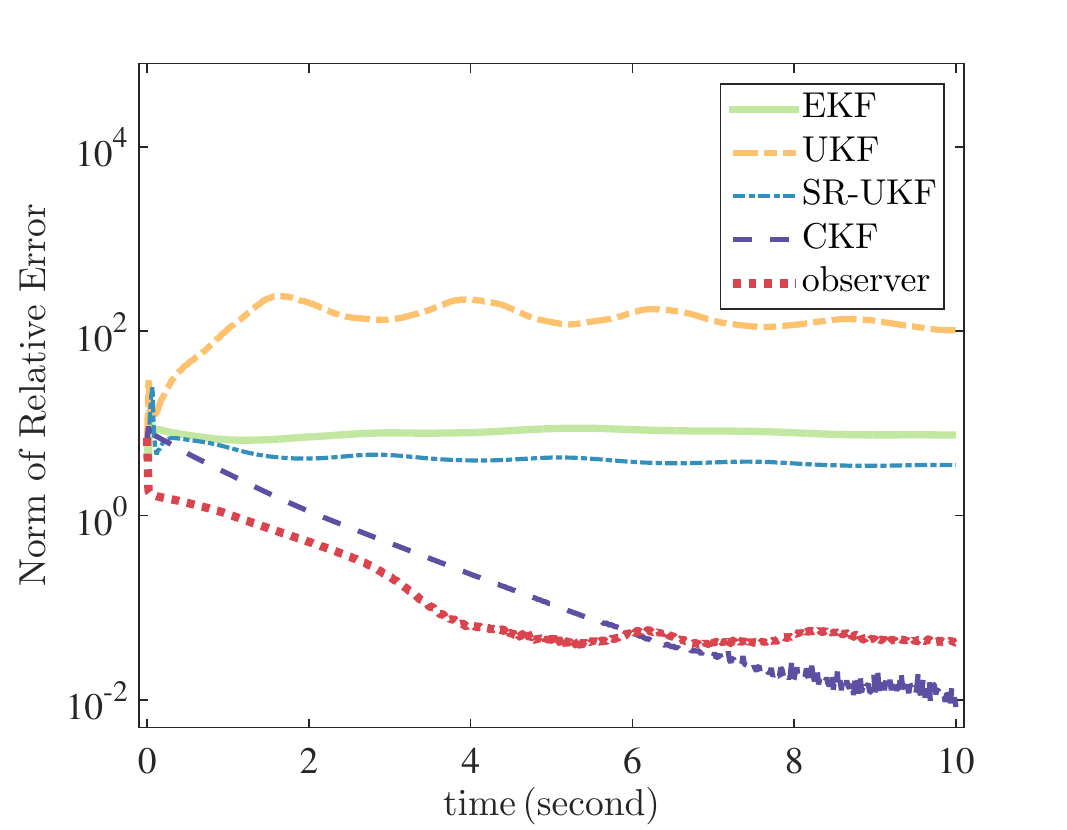}}
			\centering
			\subfloat[]{\label{sce3_norm}\includegraphics[width=1.81in]{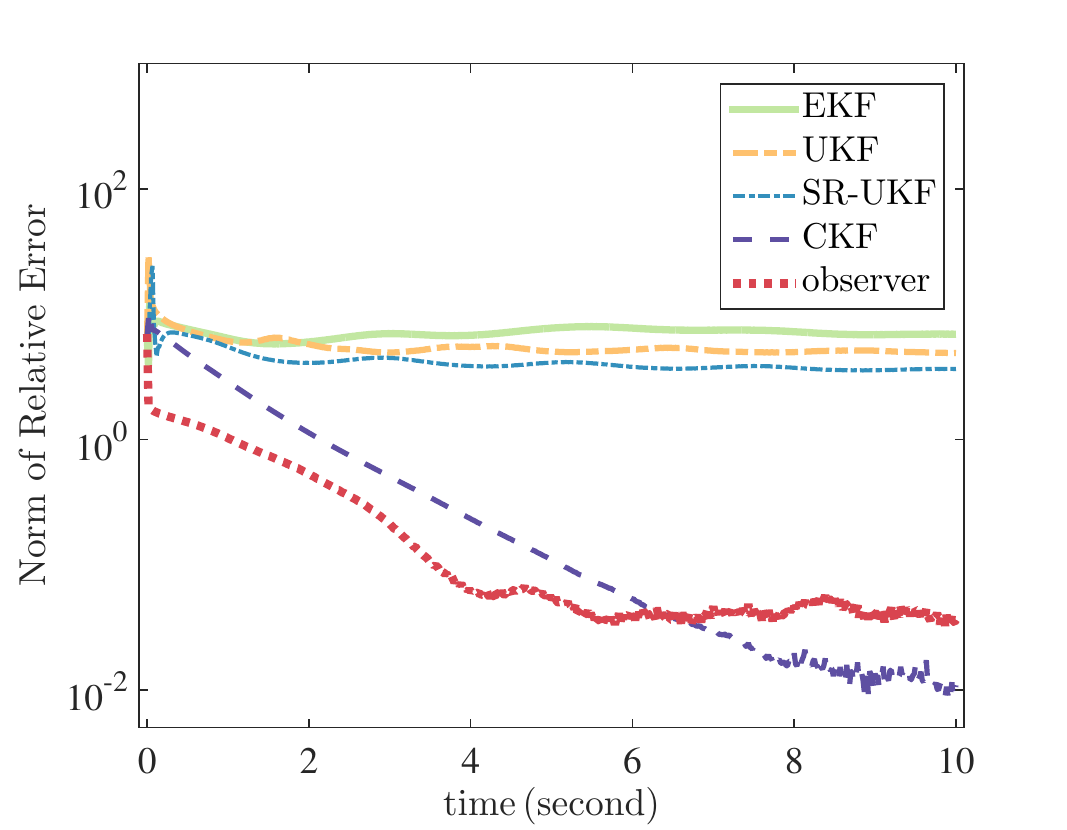}}
		\end{minipage}  
		\caption{Norm of relative error of the states. (a) Scenario 2. (b) Scenario 3.}
		\label{sce2_3_norm}
	\end{figure}

	
	

	\subsection{Discussion on Model Uncertainty Estimation}
	
	We take Scenario 1 as an example to discuss the performance of different methods in dealing with model uncertainty. 
	The states of the system with and without model uncertainty, including unknown inputs, unavailable inputs, and parameter inaccuracy, are separately denoted by $\vect{x}$ and $\vect{x}_0$, which are shown in Fig. \ref{sce1_x_x0}. 
	The difference between $\vect{x}$ and $\vect{x}_0$, $\vect{x}-\vect{x}_0$, is shown in Fig. \ref{sce1_UI}. 
	The estimated model uncertainty for Generator 1 by EKF, SR-UKF, CKF, and the observer is shown in Fig. \ref{sce1_UI_est} and that for 
	UKF is shown in Fig. \ref{sce1_UI_est_ukf}. 
	It is seen that SR-UKF, CKF, and the observer can estimate the model uncertainty pretty well while
	the EKF does not perform as well and the UKF has the worst performance for which the model uncertainty estimation is largely misled by the data integrity attack.
	
\begin{figure}[h]
	\centering
	\includegraphics[width=3.3in]{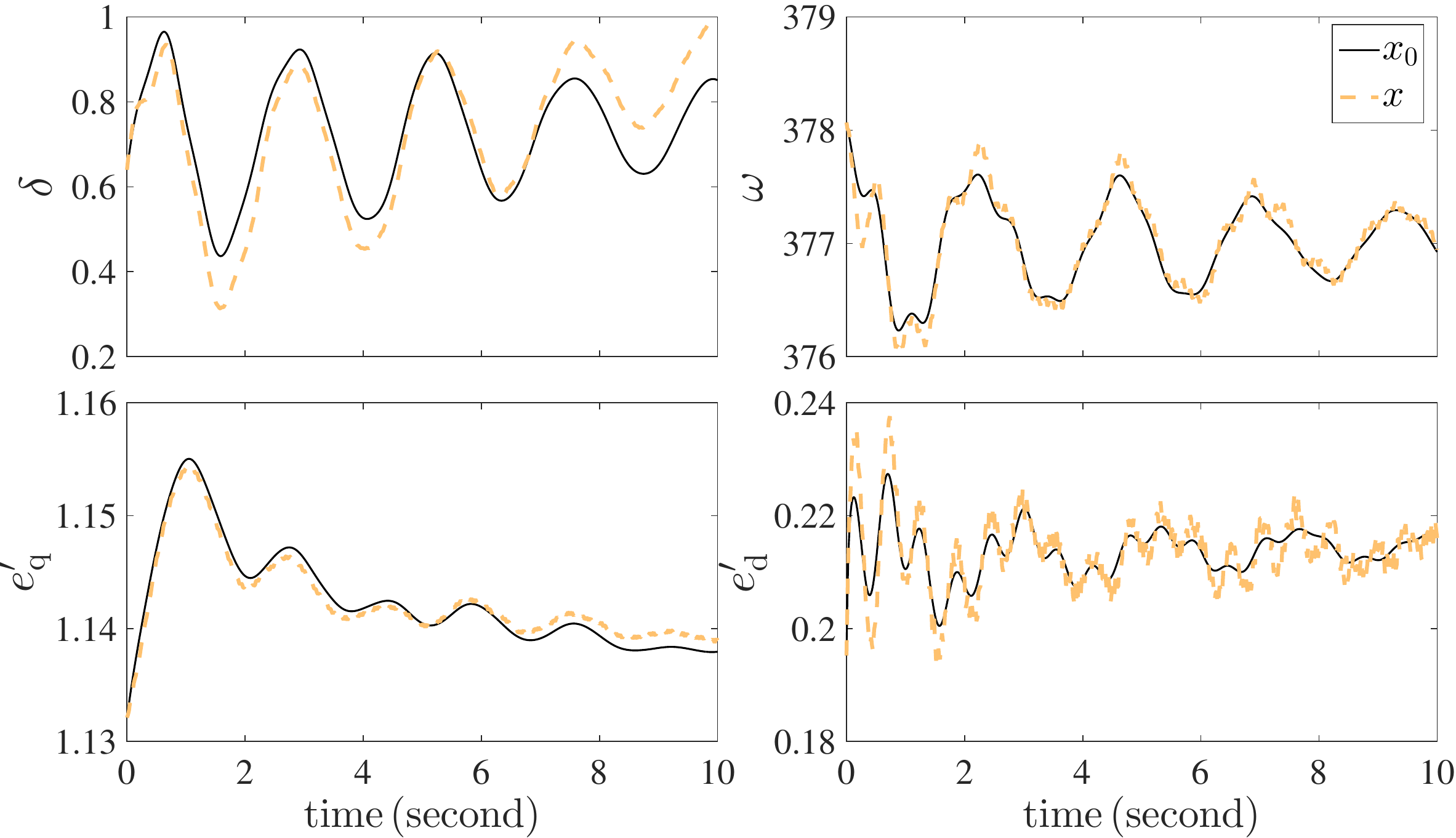}
	\caption{System states with and without model uncertainty in Scenario 1.}
	\label{sce1_x_x0}
\end{figure}

\begin{figure}[h]
	\centering
	\includegraphics[width=3.3in]{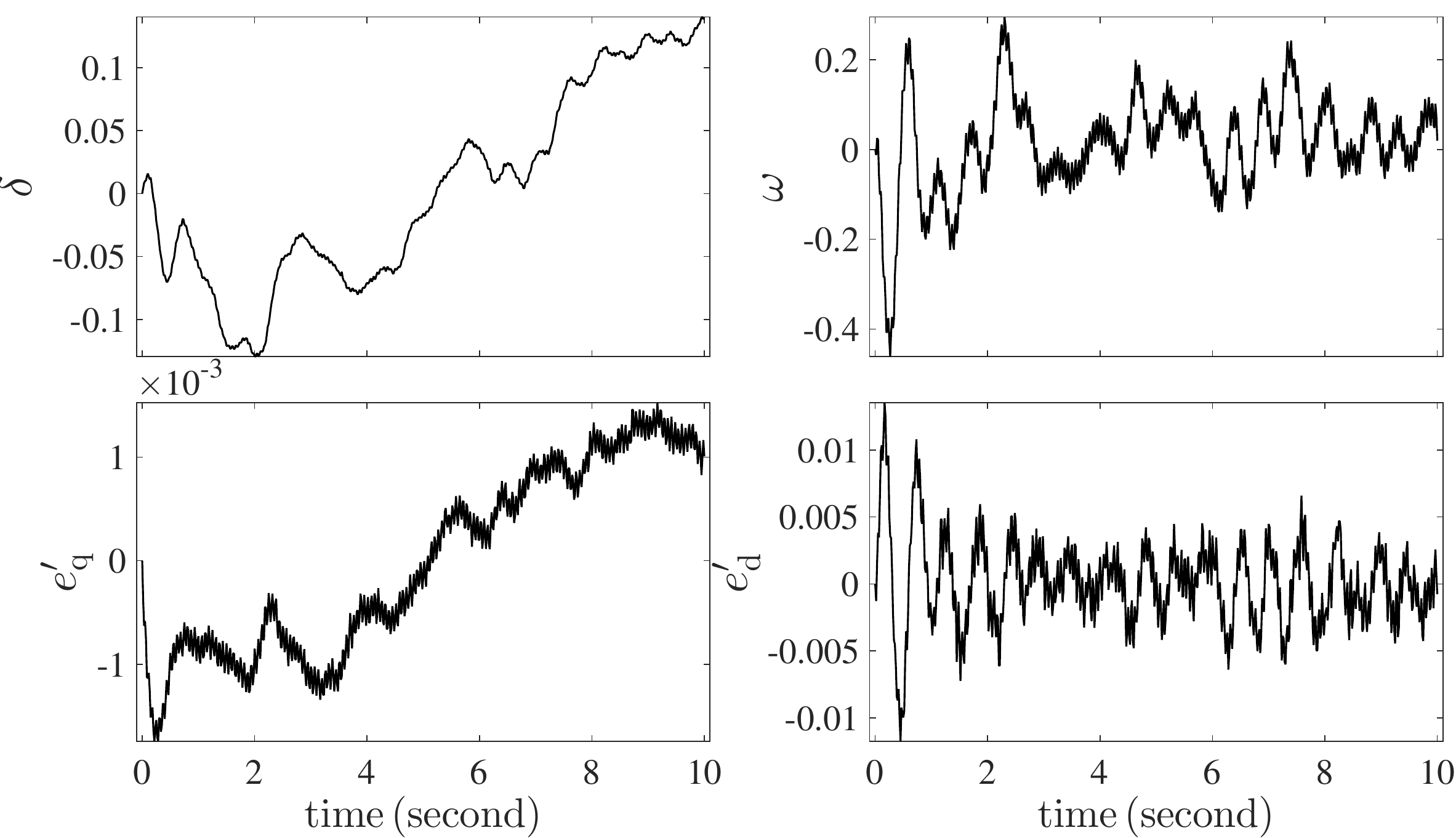}
	\caption{The $\boldsymbol{x}-\boldsymbol{x}_0$ in Scenario 1.}
	\label{sce1_UI}
\end{figure}

	\begin{figure}[!t]
		\centering
		\includegraphics[width=3.3in]{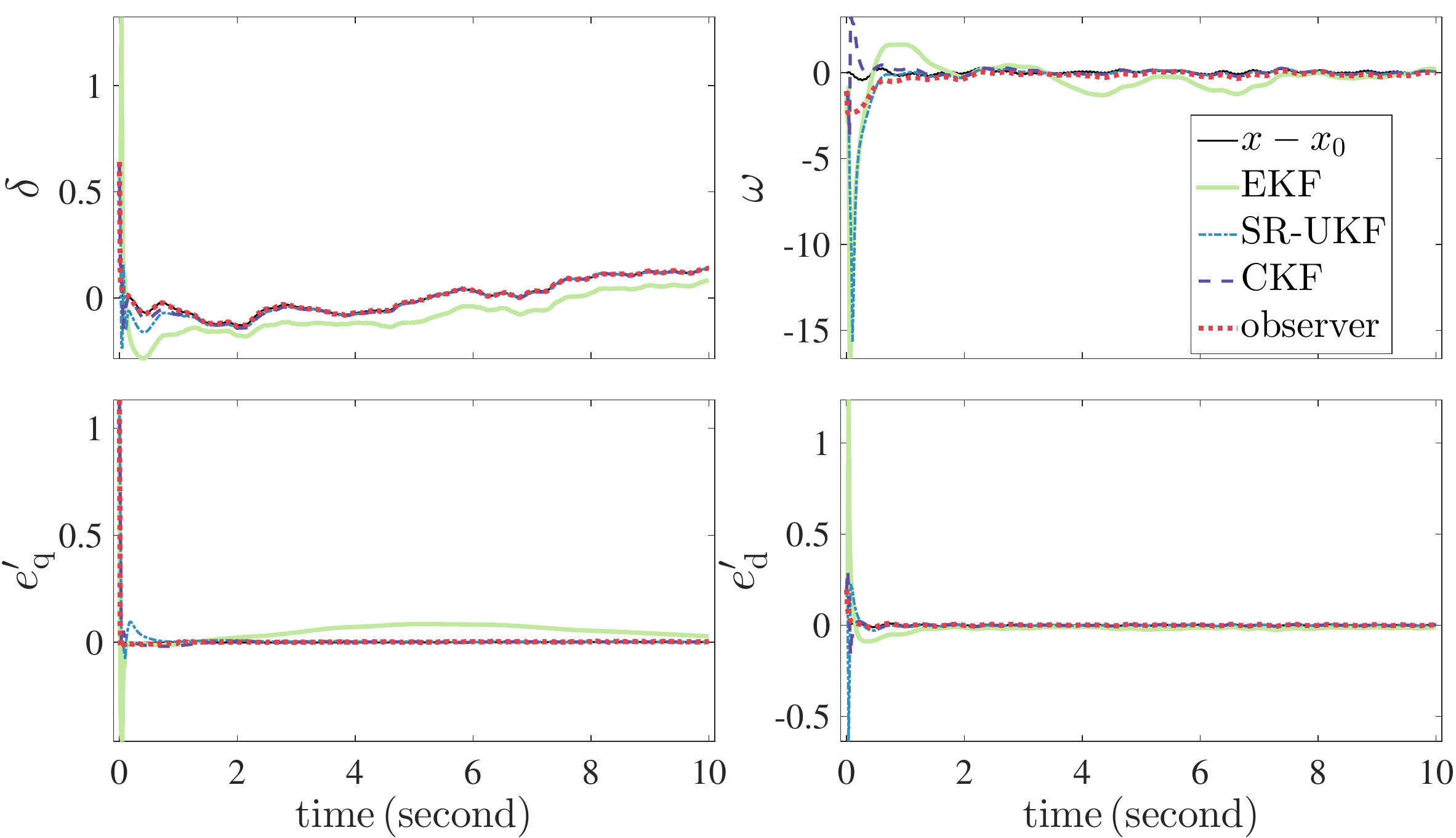}
		\caption{Estimated model uncertainty for EKF, SR-UKF, CKF, and the observer in Scenario 1.}
		\label{sce1_UI_est}
	\end{figure}

	\begin{figure}[!t]
		\centering
		\includegraphics[width=3.3in]{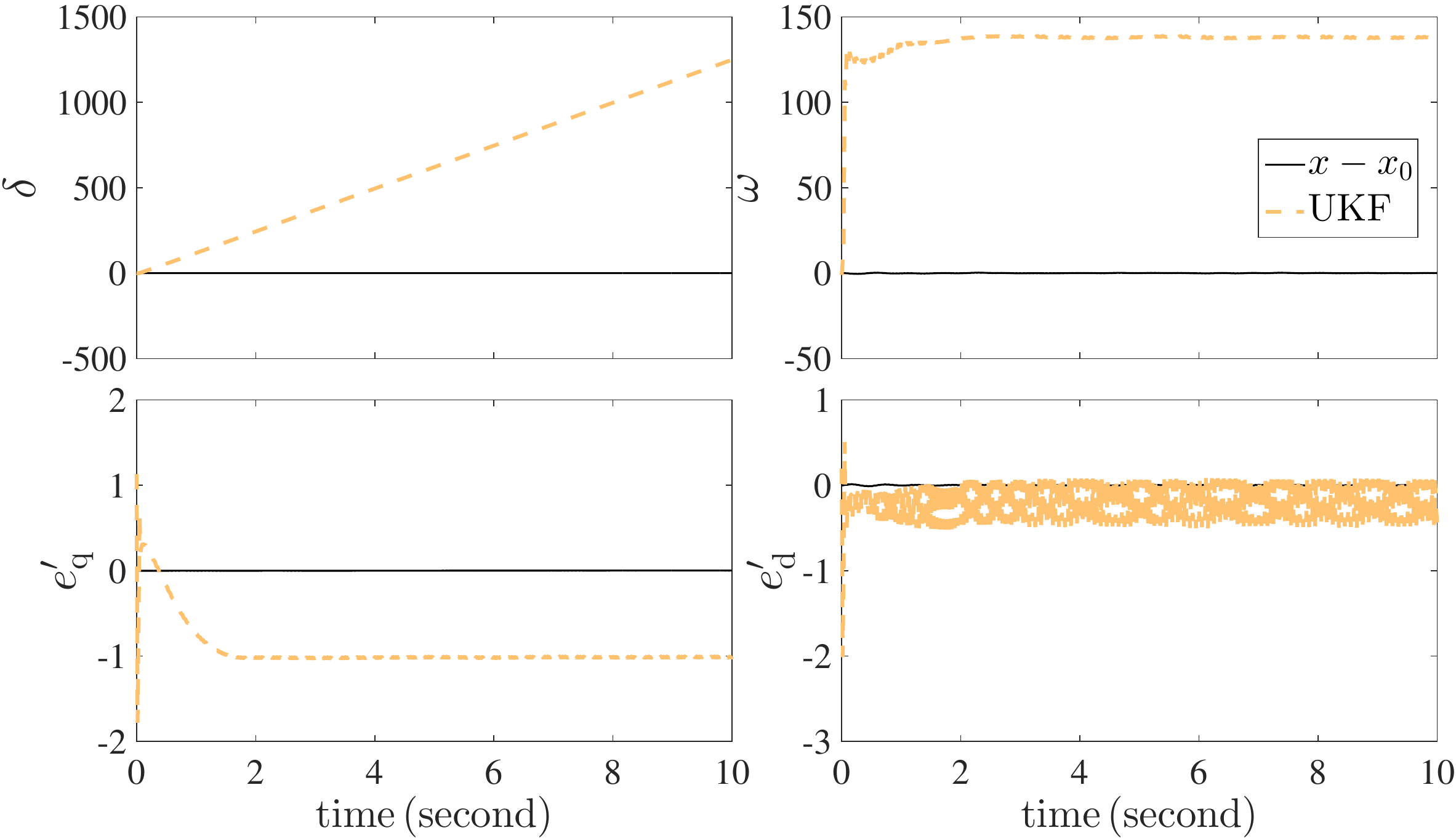}
		\caption{Estimated model uncertainty for UKF in Scenario 1.}
		\label{sce1_UI_est_ukf}
	\end{figure}

	\subsection{Discussion on Cyber Attack Detection}
	
	The normalized innovation ratio of the $j$th measurement at time step $k$ is defined as the ratio between the deviation of its actual measurement 
	from the predicted measurement and the expected standard deviation \cite{valverde2011unscented,pwukf1,pwukf2}: 
	\begin{equation}
	\lambda_{k,j}=\frac{y_{k,j}-\hat{y}_{k|k-1,j}}{\sqrt{\vect{P}_{yy,k|k-1,j}}},
	\end{equation}
	where $\vect{P}_{yy,k|k-1,j}$ is the $j$th diagonal element of the measurement covariance. 
	
	The normalized innovation ratio for all of the measurements for EKF, UKF, SR-UKF, and CKF in Scenario 1 are shown in Fig. \ref{sce1_baddata_kf}. 
	It is seen that for EKF and UKF the normalized innovation ratios of a few uncompromised measurements are greater than those for the compromised measurements, 
	which means that EKF and UKF cannot correctly detect the compromised measurements. 
	For SR-UKF and CKF, after a few seconds (in the first second some uncompromised measurements can have bigger normalized innovation ratios mainly because the parameters used for estimation in that time period are inaccurate), 
	the normalized innovation ratios for compromised measurements are significantly greater than those for the uncompromised ones, and the compromised measurements can be 
	detected by a properly chosen threshold. 
	Compared to SR-UKF, CKF has a better performance. 
	For Scenarios 2--3 the results are similar 
	and are not presented.
	
	\begin{figure}[!t]
		\begin{minipage}{.5\linewidth}
			\centering
			\subfloat[]{\label{sce1_baddata_ekf}\includegraphics[width=1.8in]{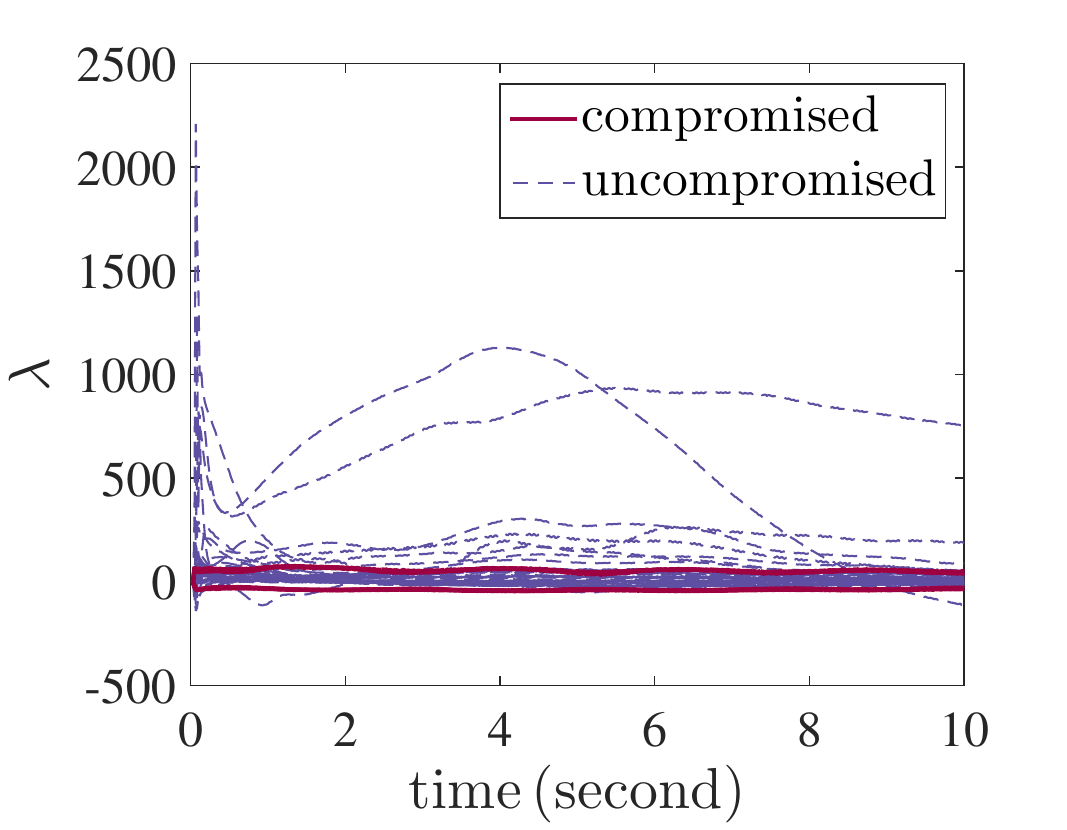}}
			\centering
			\subfloat[]{\label{sce1_baddata_ukf}\includegraphics[width=1.8in]{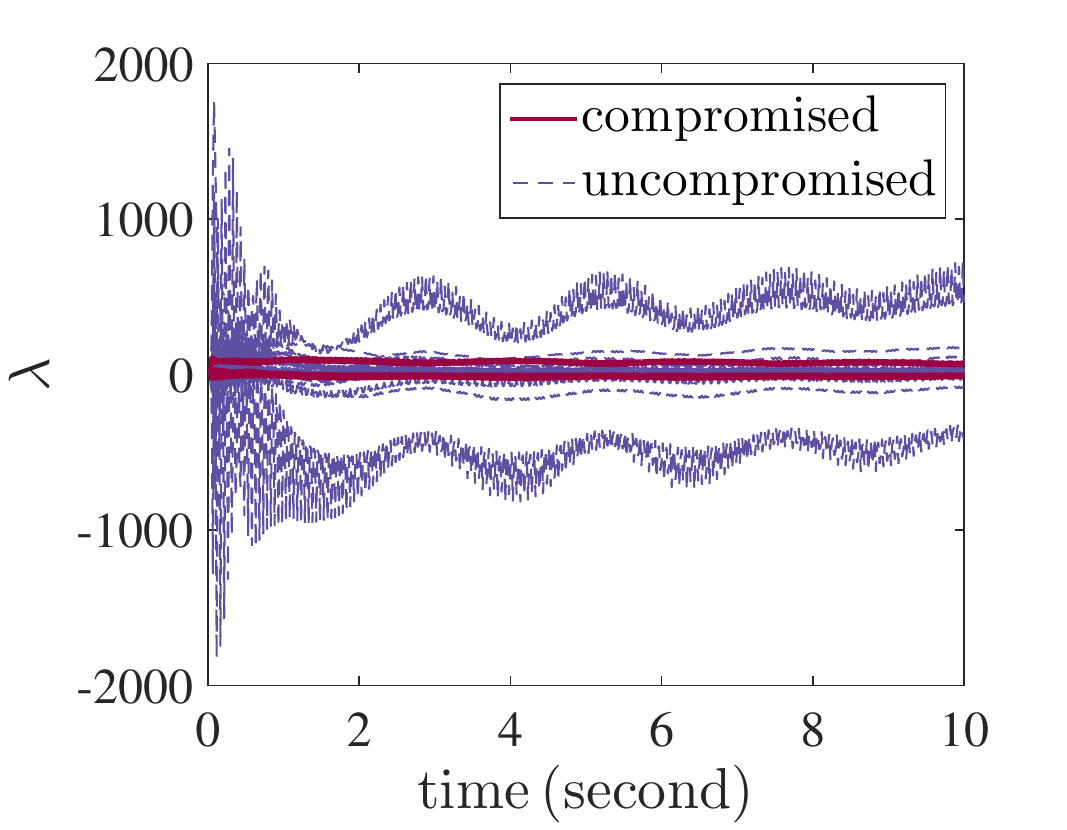}}
		\end{minipage}
		
		\vspace*{-0.4cm}
		\centering
		\subfloat[]{\label{sce1_baddata_srukf}\includegraphics[width=1.8in]{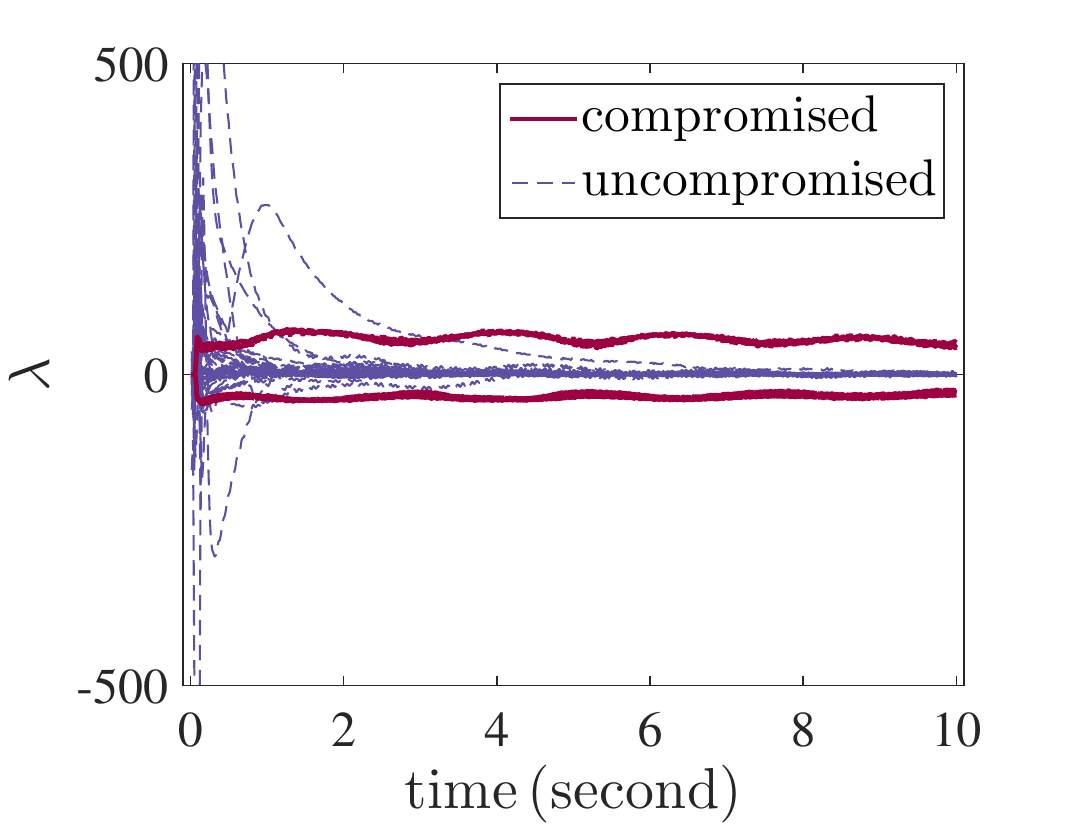}}
		\centering
		\subfloat[]{\label{sce1_baddata_ckf}\includegraphics[width=1.8in]{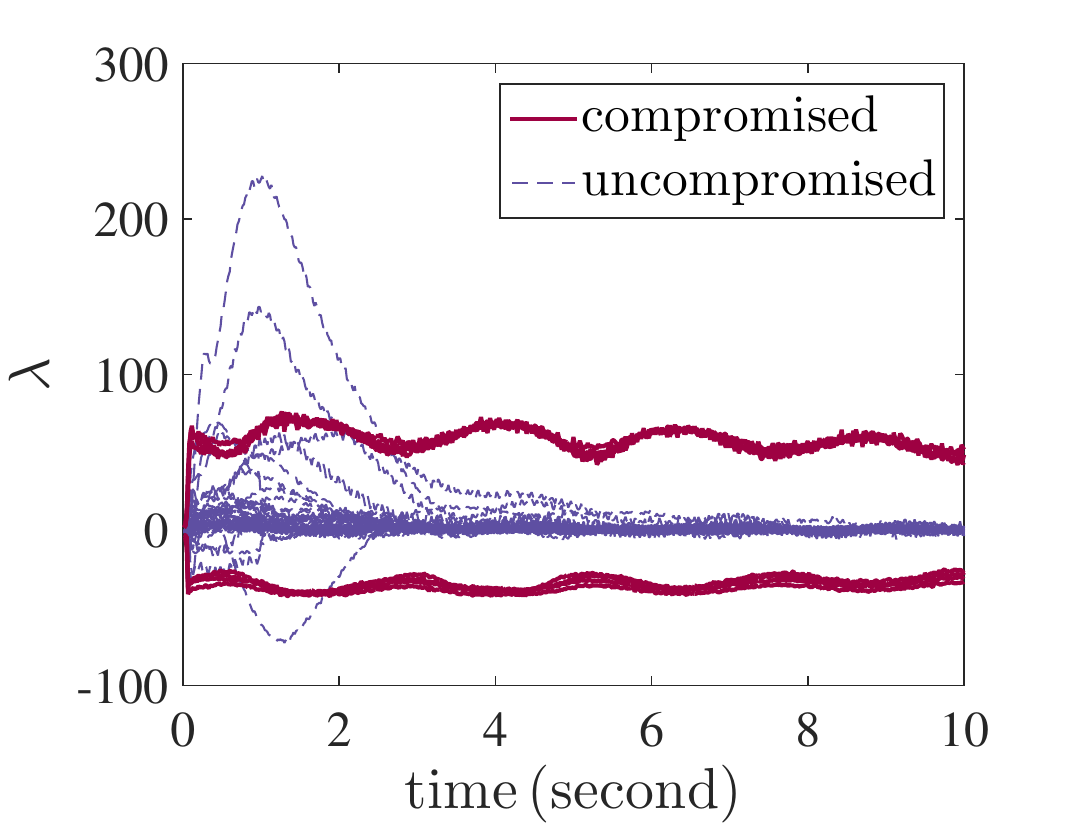}}
		
		\caption{Cyber attack detection in Scenario 1 for (a) EKF, (b) UKF, (c) SR-UKF, and (d) CKF.}
		\label{sce1_baddata_kf}
	\end{figure}

	For the observer, since there is no measurement covariance we will detect cyber attacks against the measurements directly using the measurement innovation $y_{k,j}-\hat{y}_{k|k-1,j}$, 
	which is shown in Fig. \ref{sce1_baddata_observer} for Scenario 1. 
	It is seen that after the first second in which the parameters are inaccurate the measurement innovation of the compromised measurements are significantly greater than those of the uncompromised ones and thus the compromised measurements can be easily detected. 
	In Fig. \ref{sce1_baddata_observer_real} we also show the different between the real and estimated measurements, $y_{0}-\hat{y}$. 
	For both the compromised and uncompromised measurements, the estimated measurements from the observer can almost immediately converge to the real measurements after the first second. 
	
	In Fig. \ref{sce23_baddata_observer} we show the measurement innovation of the observer for Scenario 2 (Fig. \ref{sce2_baddata_observer}) and Scenario 3 (Fig. \ref{sce3_baddata_observer}), 
	which indicates that the compromised measurements can also be detected by the observer. 

	\begin{figure}[!t]
		\begin{minipage}{.5\linewidth}
			\centering
			\subfloat[]{\label{sce1_baddata_observer}\includegraphics[width=1.81in]{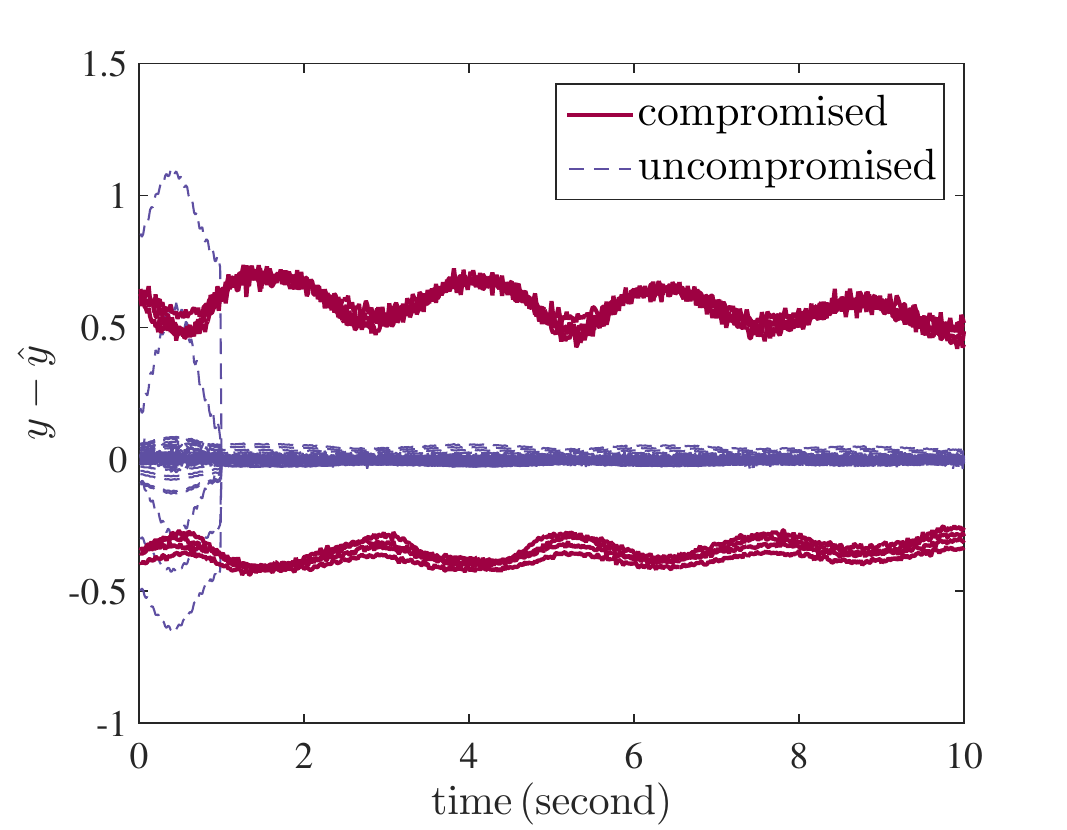}}
			\centering
			\subfloat[]{\label{sce1_baddata_observer_real}\includegraphics[width=1.81in]{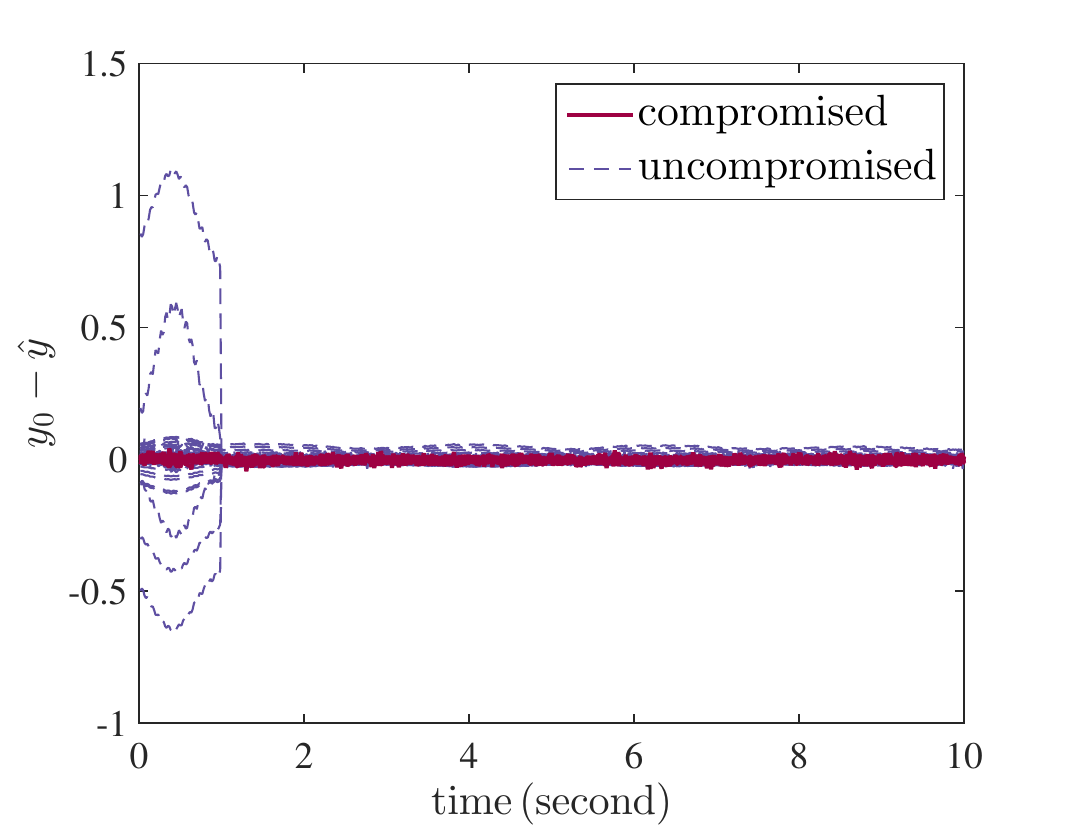}}
		\end{minipage}  
		\caption{Norm of relateive error of the states. (a) attack detection and real and estimated measurements for the observer in Scenario 1.}
		\label{sce1_baddata_observer_real_meas}
	\end{figure}

	%
	%
	%

	\begin{figure}[!t]
	\begin{minipage}{.5\linewidth}
	\centering
	\subfloat[]{\label{sce2_baddata_observer}\includegraphics[width=1.81in]{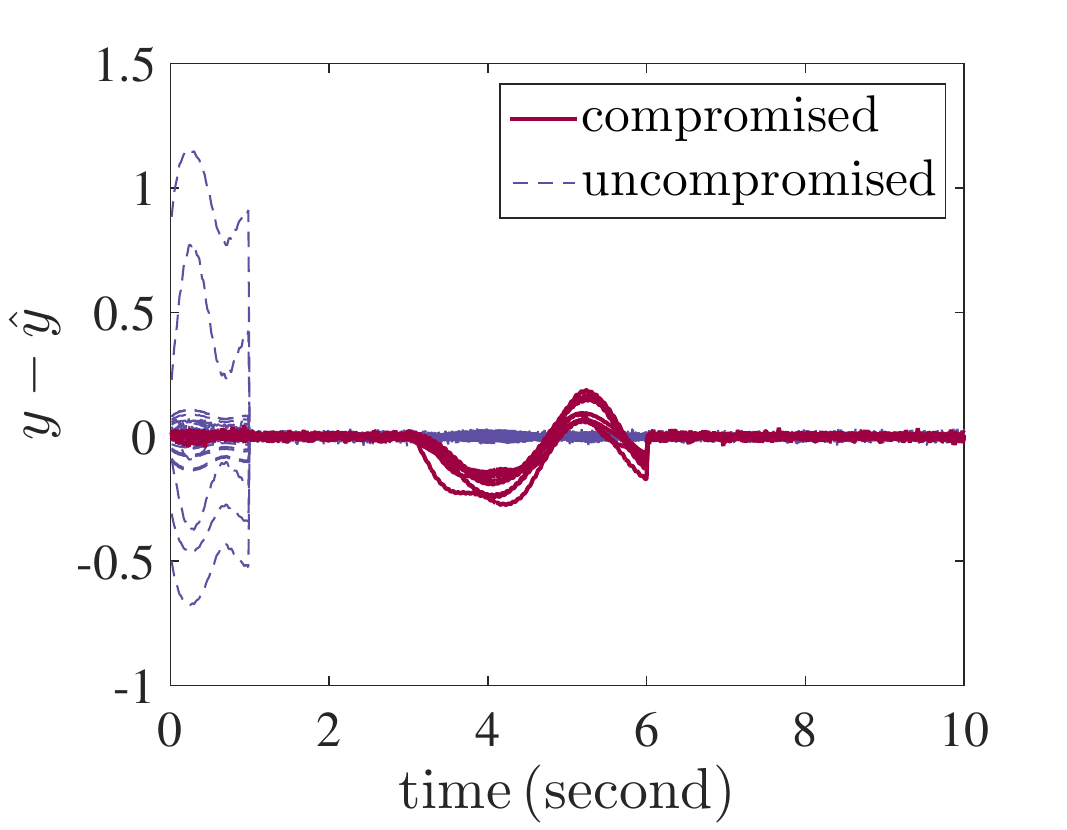}}
	\centering
	\subfloat[]{\label{sce3_baddata_observer}\includegraphics[width=1.81in]{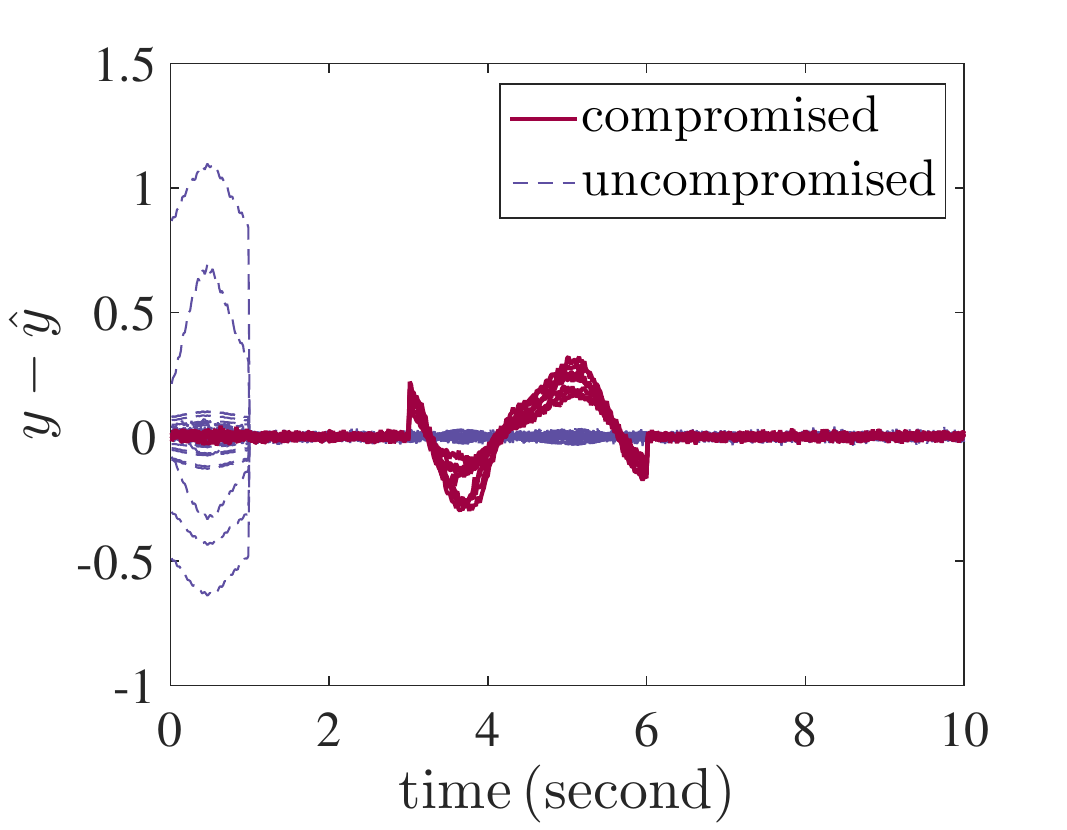}}
	\end{minipage}  
	\caption{Cyber attack detection in Scenario 2 and Scenario 3 for the observer.}
	\label{sce23_baddata_observer}
	\end{figure}

	\subsection{Non-Gaussian Measurement Noise}
	
	We performed DSE under data integrity attack in Scenario 1 with non-Gaussian measurement noise, including the Laplace noise and Cauchy noise. 
	
	Laplace noise with mean $m$ and scale $s$ is generated by
	\begin{equation}
	r_\mathrm{Laplace}=m - s \, \mathrm{sgn}(U_1)\ln (1-2|U_1|),
	\end{equation}
	where $m$ is set to be zero, $s$ is chosen as $0.02$, and $U_1$ is a random number sampled from a uniform distribution in the interval $(-0.5, 0.5]$.
	
	Cauchy noise is obtained by sampling the inverse cumulative distribution function of the distribution
	\begin{equation}
	r_\mathrm{Cauchy} = a + b \tan \big(\pi (U_2-0.5)\big),
	\end{equation}
	where $a=0$ and $b=10^{-4}$ are the location and scale parameters, and $U_2$ is randomly sampled from the uniform distribution on the
	interval $(0,1)$.
	
	The norms of the relative error of the states under Laplace and Cauchy noises are shown in Fig. \ref{sce1_laplace_cauchy}. 
	Similar to the case with Gaussian noise, the observer and CKF also outperform the other methods. 
	Under Laplace noise, the performance of different methods are similar to that under Gaussian noise. 
	However, under Cauchy noise that has a super-heavy tailed distribution with no defined moments, the performance of all methods degrade, converging to a much bigger norm of relative error of the states. 
	
	\begin{figure}[!t]
	\begin{minipage}{.5\linewidth}
		\centering
		\subfloat[]{\label{sce1_laplace}\includegraphics[width=1.8in]{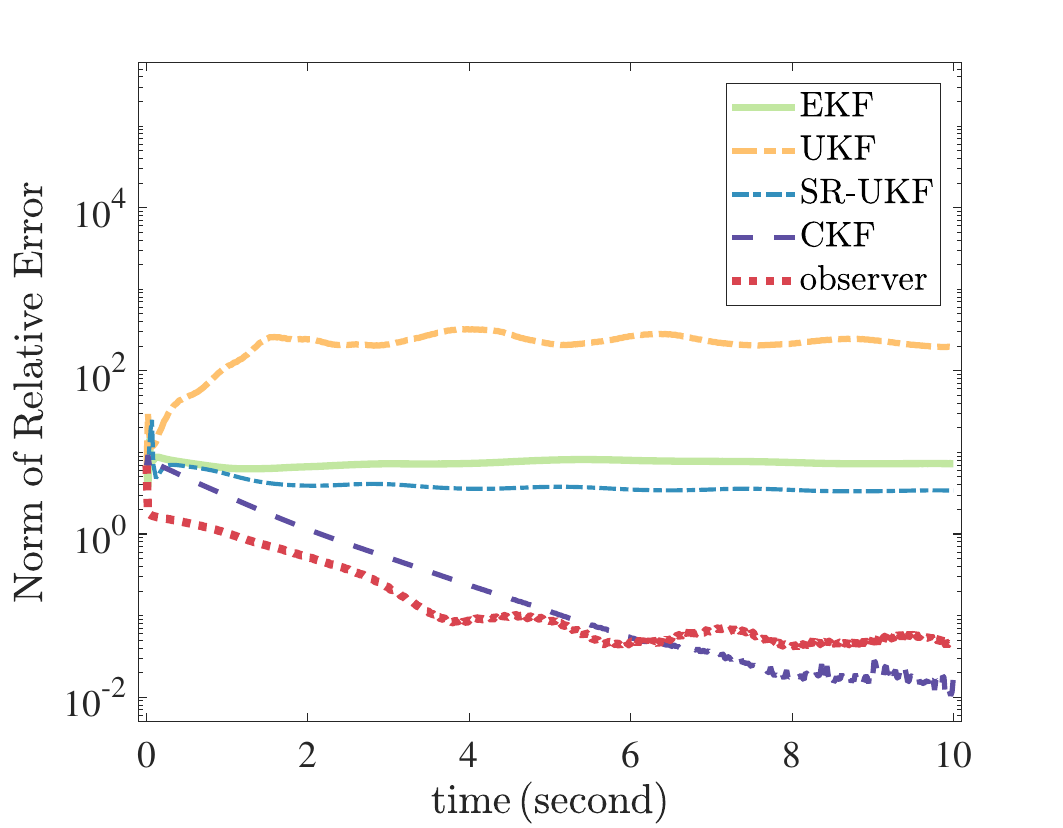}}
		\centering
		\subfloat[]{\label{sce1_cauchy}\includegraphics[width=1.8in]{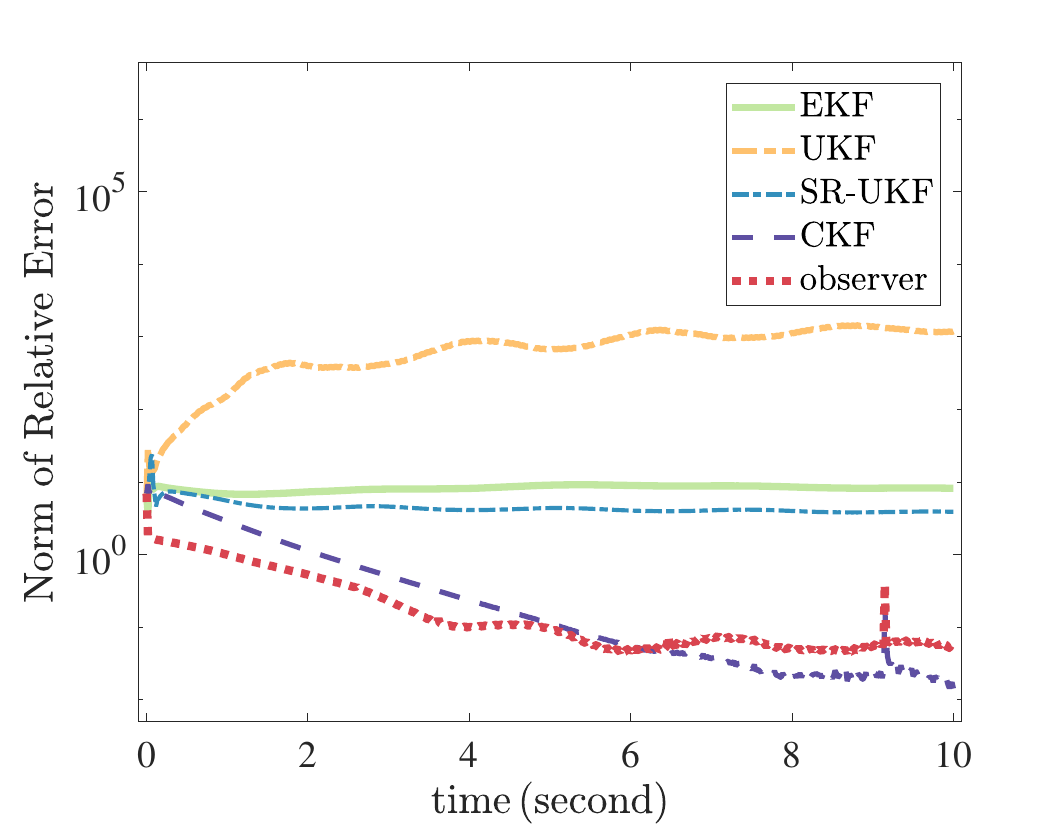}}
	\end{minipage}  
	\caption{Norm of relative error of the states under different measurement noises. (a) Laplace noise and (b) Cauchy noise.}
	\label{sce1_laplace_cauchy}
\end{figure}

	\subsection{Computational Efficiency}
	
	For the above three scenarios, the time for estimation by different methods is listed in Table \ref{time}. 
	It is seen that EKF and the observer are more efficient than the other methods while CKF is the least efficient.
	Note that the time reported here is from MATLAB implementations. 
	It can be greatly reduced by more efficient, such as C-based implementations. 
	
	\begin{table}[!t]
		\footnotesize
		\renewcommand{\arraystretch}{1.1}
		\caption{Time for Performing Estimation for 10 Seconds}
		\label{time}
		\centering 
		\begin{tabular}{ccccc}
			\hline
			EKF & UKF & SR-UKF & CKF & observer \\
			\midrule
			$4.0\,\textrm{s}$ & $11.2\,\textrm{s}$ & $11.6\,\textrm{s}$    &   $9.9\,\textrm{s}$  & $5.8\,\textrm{s}$  \\
			\hline
		\end{tabular}
	\end{table}

	\section{Comparing Kalman Filters and Observers} \label{sec:remarks}
	
	Here, various functionalities of DSE methods and their strengths and weaknesses relative to each functionality are presented based on 
	(a) the technical, theoretical capabilities and (b) experimental results in Section~\ref{sec:NumericalResults}. 
	
	\begin{itemize}\setlength\itemsep{0.2em}
		\item \textit{Nonlinearities in Dynamics:} UKF, SR-UKF, CKF, and the observer in Section \ref{sec:Observers} all work on nonlinear systems  
		while EKF assumes linearized system dynamics. Besides, the presented observer 
		uses linearized measurement functions for design but directly uses nonlinear measurement functions for estimation. 
		
		\item \textit{Solution Feasibility:} The main principle that governs the design of most observers is based on finding a matrix gain satisfying a certain condition, such as a solution to a matrix inequality. 
		The state estimates are guaranteed to converge to the actual ones if a solution to the LMI exists. In contrast, KF methods do not require that.
		
		\item \textit{Unknown Initial Conditions:} Most observer designs are independent on the knowledge of the initial conditions of the system. However, if the estimator's initial condition is chosen to be reasonably different from the actual one, estimates from KF might not converge to the actual ones.
		
		\item \textit{Robustness to Model Uncertainty and Cyber Attacks:} The observer in Section \ref{sec:Observers} and the CKF outperforms UKF (SR-UKF) and EKF in the state estimation under model uncertainty and attack vectors. The observer is robust to model uncertainties because it only assumes that the nonlinearities in the power system dynamics (i.e., $\boldsymbol{\phi}(\vect{x})$) satisfy the quadratic inner-boundedness and the one-sided Lipschitz condition. As in Section \ref{sec:CKFF}, CKF is more robust mostly due to its more accurate cubature approach, which, however, requires more careful investigation. 
		
		\item \textit{Tolerance to Process and Measurement Noise:} The observer in Section \ref{sec:Observers} is tolerant to measurement and process noise similar to those assumed for KFs. By design, the KF techniques are developed to deal with such noise. 
		
		\item \textit{Convergence Guarantees:} Observers have theoretical guarantees for convergence while for the KF techniques there is no strict proof to guarantee that the estimation converges to actual states.
		
		\item  \textit{Numerical Stability:} Observers do not have numerical stability problems while UKF can encounter numerical instability because the estimation error covariance matrix is not always guaranteed to be positive semi-definite \cite{qi2015dynamic}.
		
		\item \textit{Tolerance to Parametric Inaccuracy:} KF-based methods can tolerate inaccurate parameters to some extent. Dynamic observers deal with parametric uncertainty in the sense that all uncertainties can be augmented to the unknown input component in the state dynamics ($\m{B}_\mathrm{w}\vect {w}$).  
		
		\item \textit{Computational Complexity:} The CKF, UKF (SR-UKF), and EKF all have computational complexity of $\mathcal{O}(n^3)$ \cite{van2001square,CKF}. 
		Since the observers' matrix gains are obtained offline by solving LMIs, observers are easier to implement as only the dynamics are needed in the estimation.
	\end{itemize}

	\section{Conclusion and Future Work}~\label{sec:conc}
	
	In this paper, we discuss different DSE methods by presenting an overview of state-of-the-art estimation techniques and developing alternatives, including the CKF and dynamic observers, to address major limitations of existing methods such as intolerance to inaccurate system model and malicious cyber attacks. 
	The proposed methods are extensively tested on a 16-machine 68-bus power system, under significant model uncertainty and cyber attacks against the synchrophaosr measurements. 
	It is shown that the CKF and the observer are more robust to model uncertainty and cyber attacks. 
	
	Based on the theoretical capabilities and the experimental results, 
	we summarize the strengths and weaknesses of different estimation techniques especially for power system DSE. 
	We acknowledge that some of these comparisons, such as tolerance to process and measurement noise, are mostly based on numerical results. As future work we will more theoretically investigate and analyze the observer in comparison with Kalman filters. 

%
%
	
	\bibliographystyle{IEEEtran}

\begin{thebibliography}{10}
\providecommand{\url}[1]{#1}
\csname url@samestyle\endcsname
\providecommand{\newblock}{\relax}
\providecommand{\bibinfo}[2]{#2}
\providecommand{\BIBentrySTDinterwordspacing}{\spaceskip=0pt\relax}
\providecommand{\BIBentryALTinterwordstretchfactor}{4}
\providecommand{\BIBentryALTinterwordspacing}{\spaceskip=\fontdimen2\font plus
\BIBentryALTinterwordstretchfactor\fontdimen3\font minus
  \fontdimen4\font\relax}
\providecommand{\BIBforeignlanguage}[2]{{%
\expandafter\ifx\csname l@#1\endcsname\relax
\typeout{** WARNING: IEEEtran.bst: No hyphenation pattern has been}%
\typeout{** loaded for the language `#1'. Using the pattern for}%
\typeout{** the default language instead.}%
\else
\language=\csname l@#1\endcsname
\fi
#2}}
\providecommand{\BIBdecl}{\relax}
\BIBdecl

\bibitem{se3}
A.~Monticelli, ``Electric power system state estimation,'' \emph{Proc. IEEE},
  vol.~88, no.~2, pp. 262--282, Feb. 2000.

\bibitem{se2}
A.~Abur and A.~Exp{\'o}sito, \emph{Power System State Estimation: Theory and
  Implementation}, ser. Power Engineering (Willis).\hskip 1em plus 0.5em minus
  0.4em\relax CRC Press, 2004.

\bibitem{se5}
G.~He, S.~Dong, J.~Qi, and Y.~Wang, ``Robust state estimator based on maximum
  normal measurement rate,'' \emph{IEEE Trans. Power Syst.}, vol.~26, no.~4,
  pp. 2058--2065, Nov. 2011.

\bibitem{qi2012power}
J.~Qi, G.~He, S.~Mei, and F.~Liu, ``Power system set membership state
  estimation,'' in \emph{IEEE Power and Energy Society General Meeting}, Jul.
  2012, pp. 1--7.

\bibitem{zhou1996robust}
K.~Zhou, J.~C. Doyle, and K.~Glover, \emph{Robust and Optimal Control}.\hskip
  1em plus 0.5em minus 0.4em\relax Prentice hall New Jersey, 1996.

\bibitem{huang2013generator}
Z.~Huang, P.~Du, D.~Kosterev, and S.~Yang, ``Generator dynamic model validation
  and parameter calibration using phasor measurements at the point of
  connection,'' \emph{IEEE Trans. Power Syst.}, vol.~28, no.~2, pp. 1939--1949,
  May 2013.

\bibitem{ariff2015estimating}
M.~Ariff, B.~Pal, and A.~Singh, ``Estimating dynamic model parameters for
  adaptive protection and control in power system,'' \emph{IEEE Trans. Power
  Syst.}, vol.~30, no.~2, pp. 829--839, Mar. 2015.

\bibitem{liu2011false}
Y.~Liu, P.~Ning, and M.~K. Reiter, ``False data injection attacks against state
  estimation in electric power grids,'' \emph{ACM Trans. Inf. Syst. Secur.},
  vol.~14, no.~1, pp. 13:1--13:33, Jun. 2011.

\bibitem{6032057}
R.~J.~T. O.~Kosut, L.~Jia and L.~Tong, ``Malicious data attacks on the smart
  grid,'' \emph{IEEE Trans. Smart Grid}, vol.~2, no.~4, pp. 645--658, Dec.
  2011.

\bibitem{6194239}
G.~D. O.~Vukovic, K. C.~Sou and H.~Sandberg, ``Network-aware mitigation of data
  integrity attacks on power system state estimation,'' \emph{IEEE J. Sel. Area
  Comm.}, vol.~30, no.~6, pp. 1108--1118, Jul. 2012.

\bibitem{6490324}
W.~Y. D. A. N. Z. W.~Z. Q.~Yang, J.~Yang, ``On false data-injection attacks
  against power system state estimation: modeling and countermeasures,''
  \emph{IEEE Tran. Parallel Distrib. Syst.}, vol.~25, no.~3, pp. 717--729, Mar.
  2014.

\bibitem{kalman1960new}
R.~E. Kalman, ``A new approach to linear filtering and prediction problems,''
  \emph{J. Fluids Eng.}, vol.~82, no.~1, pp. 35--45, Mar. 1960.

\bibitem{jazwinski2007stochastic}
A.~H. Jazwinski, \emph{Stochastic Processes and Filtering Theory}.\hskip 1em
  plus 0.5em minus 0.4em\relax Courier Corporation, 2007.

\bibitem{ukf}
S.~J. Julier and J.~K. Uhlmann, ``New extension of the {K}alman filter to
  nonlinear systems,'' in \emph{AeroSense'97}.\hskip 1em plus 0.5em minus
  0.4em\relax International Society for Optics and Photonics, Jul. 1997, pp.
  182--193.

\bibitem{ukf1}
S.~Julier and J.~Uhlmann, ``Unscented filtering and nonlinear estimation,''
  \emph{Proc. IEEE}, vol.~92, no.~3, pp. 401--422, Mar. 2004.

\bibitem{CKF}
I.~Arasaratnam and S.~Haykin, ``Cubature {K}alman filters,'' \emph{IEEE Trans.
  Autom. Control}, vol.~54, no.~6, pp. 1254--1269, Jun. 2009.

\bibitem{kang2013survey}
W.~Kang, A.~J. Krener, M.~Xiao, and L.~Xu, ``A survey of observers for
  nonlinear dynamical systems,'' in \emph{Data Assimilation for Atmospheric,
  Oceanic and Hydrologic Applications (Vol. II)}.\hskip 1em plus 0.5em minus
  0.4em\relax Springer, 2013, pp. 1--25.

\bibitem{Radke2006}
A.~Radke and Z.~Gao, ``A survey of state and disturbance observers for
  practitioners,'' in \emph{American Control Conference}, Jun. 2006, pp.
  5183--5188.

\bibitem{Hidayat2011}
Z.~Hidayat, R.~Babuska, B.~D. Schutter, and A.~Núñez, ``Observers for linear
  distributed-parameter systems: A survey,'' in \emph{IEEE International
  Symposium on Robotic and Sensors Environments (ROSE)}, Sept. 2011, pp.
  166--171.

\bibitem{ekf1}
Z.~Huang, K.~Schneider, and J.~Nieplocha, ``Feasibility studies of applying
  {K}alman filter techniques to power system dynamic state estimation,'' in
  \emph{Proc. 2007 Power Engineering Conf.}, Dec. 2007, pp. 376--382.

\bibitem{ekf2}
E.~Ghahremani and I.~Kamwa, ``Dynamic state estimation in power system by
  applying the extended {K}alman filter with unknown inputs to phasor
  measurements,'' \emph{IEEE Trans. Power Syst.}, vol.~26, no.~4, pp.
  2556--2566, Nov. 2011.

\bibitem{valverde2011unscented}
G.~Valverde and V.~Terzija, ``Unscented {K}alman filter for power system
  dynamic state estimation,'' \emph{IET Gener. Transm. Distrib.}, vol.~5,
  no.~1, pp. 29--37, Jan. 2011.

\bibitem{ghahremani2011online}
E.~Ghahremani and I.~Kamwa, ``Online state estimation of a synchronous
  generator using unscented {K}alman filter from phasor measurements units,''
  \emph{IEEE Trans. Energy Convers.}, vol.~26, no.~4, pp. 1099--1108, Dec.
  2011.

\bibitem{pwukf1}
S.~Wang, W.~Gao, and A.~Meliopoulos, ``An alternative method for power system
  dynamic state estimation based on unscented transform,'' \emph{IEEE Trans.
  Power Syst.}, vol.~27, no.~2, pp. 942--950, May 2012.

\bibitem{pwukf2}
A.~Singh and B.~Pal, ``Decentralized dynamic state estimation in power systems
  using unscented transformation,'' \emph{IEEE Trans. Power Syst.}, vol.~29,
  no.~2, pp. 794--804, Mar. 2014.

\bibitem{sun2016power}
K.~Sun, J.~Qi, and W.~Kang, ``Power system observability and dynamic state
  estimation for stability monitoring using synchrophasor measurements,''
  \emph{Control Engineering Practice}, vol.~53, pp. 160--172, Aug. 2016.

\bibitem{van2001square}
R.~Van Der~Merwe and E.~A. Wan, ``The square-root unscented {K}alman filter for
  state and parameter-estimation,'' in \emph{Proc. Int. Conf. Acoustics,
  Speech, and Signal Processing}, vol.~6, May 2001, pp. 3461--3464.

\bibitem{qi}
J.~Qi, K.~Sun, and W.~Kang, ``Optimal {PMU} placement for power system dynamic
  state estimation by using empirical observability gramian,'' \emph{IEEE
  Trans. Power Syst.}, vol.~30, no.~4, pp. 2041--2054, Jul. 2015.

\bibitem{qi2015dynamic}
J.~Qi, K.~Sun, J.~Wang, and H.~Liu, ``Dynamic state estimation for
  multi-machine power system by unscented {K}alman filter with enhanced
  numerical stability,'' \emph{IEEE Trans. Smart Grid}, vol.~9, no.~2, pp.
  1184--1196, Mar. 2018.

\bibitem{zhou}
N.~Zhou, D.~Meng, and S.~Lu, ``Estimation of the dynamic states of synchronous
  machines using an extended particle filter,'' \emph{IEEE Trans. Power Syst.},
  vol.~28, no.~4, pp. 4152--4161, Nov. 2013.

\bibitem{cui2015particle}
Y.~Cui and R.~Kavasseri, ``A particle filter for dynamic state estimation in
  multi-machine systems with detailed models,'' \emph{IEEE Trans. Power Syst.},
  vol.~30, no.~6, pp. 3377--3385, Nov. 2015.

\bibitem{zhou2015dynamic}
N.~Zhou, D.~Meng, Z.~Huang, and G.~Welch, ``Dynamic state estimation of a
  synchronous machine using pmu data: A comparative study,'' \emph{IEEE Trans.
  Smart Grid}, vol.~6, no.~1, pp. 450--460, Jan. 2015.

\bibitem{gandhi2010robust}
M.~A. Gandhi and L.~Mili, ``Robust kalman filter based on a generalized
  maximum-likelihood-type estimator,'' \emph{IEEE Trans. Signal Process.},
  vol.~58, no.~5, pp. 2509--2520, May 2010.

\bibitem{zhang2014two}
G.~B. J.~Zhang, G.~Welch and Z.~Huang, ``A two-stage kalman filter approach for
  robust and real-time power system state estimation,'' \emph{IEEE Trans.
  Sustain. Energy}, vol.~5, no.~2, pp. 629--636, Apr. 2014.

\bibitem{zhao1}
J.~Zhao, M.~Netto, and L.~Mili, ``A robust iterated extended kalman filter for
  power system dynamic state estimation,'' \emph{IEEE Trans. Power Syst.},
  vol.~32, no.~4, pp. 3205--3216, July 2017.

\bibitem{zhao2}
J.~Zhao, ``Dynamic state estimation with model uncertainties using ${H}_\infty$
  extended {K}alman filter,'' \emph{IEEE Trans. Power Syst.}, vol.~33, no.~1,
  pp. 1099--1100, Jan 2018.

\bibitem{zhao3}
J.~Zhao and L.~Mili, ``Robust unscented kalman filter for power system dynamic
  state estimation with unknown noise statistics,'' \emph{IEEE Trans. Smart
  Grid}, pp. 1--1, 2017.

\bibitem{7491374}
J.~W. A.~F.~Taha, J.~Qi and J.~H. Panchal, ``Risk mitigation for dynamic state
  estimation against cyber attacks and unknown inputs,'' \emph{IEEE Trans.
  Smart Grid}, vol.~9, no.~2, pp. 886--899, Mar. 2018.

\bibitem{hajnoroozi2015generating}
A.~Hajnoroozi, F.~Aminifar, H.~Ayoubzadeh \emph{et~al.}, ``Generating unit
  model validation and calibration through synchrophasor measurements,''
  \emph{IEEE Trans. Smart Grid}, vol.~6, no.~1, pp. 441--449, Jan. 2015.

\bibitem{Chen2012}
J.~Chen and R.~Patton, \emph{Robust Model-Based Fault Diagnosis for Dynamic
  Systems}.\hskip 1em plus 0.5em minus 0.4em\relax Springer Publishing Company,
  Incorporated, 2012.

\bibitem{Pertew2005}
A.~Pertew, H.~Marquezz, and Q.~Zhao, ``Design of unknown input observers for
  lipschitz nonlinear systems,'' in \emph{Proc. American Control Conf.}, Jun.
  2005, pp. 4198--4203.

\bibitem{NESCOR2014}
``{Electric Sector Failure Scenarios and Impact Analyses},'' Electric Power
  Research Institute (EPRI), Tech. Rep., Jun. 2014.

\bibitem{sridhar2012cyber}
S.~Sridhar, A.~Hahn, and M.~Govindarasu, ``Cyber--physical system security for
  the electric power grid,'' \emph{Proc. IEEE}, vol. 100, no.~1, pp. 210--224,
  Jan. 2012.

\bibitem{error}
S.~Wang, J.~Zhao, Z.~Huang, and R.~Diao, ``Assessing {G}aussian assumption of
  pmu measurement error using field data,'' \emph{IEEE Trans. Power Del.},
  vol.~PP, no.~99, pp. 1--1, 2017.

\bibitem{ut}
J.~K. Uhlmann, ``Simultaneous map building and localization for real time
  applications,'' \emph{transfer thesis, Univ. Oxford, Oxford, UK}, 1994.

\bibitem{bellman}
R.~Bellman and R.~Bellman, \emph{Adaptive Control Processes: A Guided Tour},
  ser. 'Rand Corporation. Research studies.\hskip 1em plus 0.5em minus
  0.4em\relax Princeton University Press, 1961.

\bibitem{gadsden2014combined}
I.~A. S.~A.~Gadsden, M. Al-Shabi and S.~R. Habibi, ``Combined cubature kalman
  and smooth variable structure filtering: A robust nonlinear estimation
  strategy,'' \emph{Signal Processing}, vol.~96, pp. 290--299, Mar. 2014.

\bibitem{Zhang2012}
W.~Zhang, H.~Su, H.~Wang, and Z.~Han, ``Full-order and reduced-order observers
  for one-sided lipschitz nonlinear systems using riccati equations,''
  \emph{Commun. Nonlinear Sci. Numer. Simul.}, vol.~17, no.~12, pp. 4968--4977,
  Dec. 2012.

\bibitem{Siljak2002}
D.~Siljak, D.~Stipanovic, and A.~Zecevic, ``Robust decentralized
  turbine/governor control using linear matrix inequalities,'' \emph{IEEE
  Trans. Power Syst.}, vol.~17, no.~3, pp. 715--722, Aug. 2002.

\bibitem{qi2017nonlinear}
J.~Qi, J.~Wang, H.~Liu, and A.~D. Dimitrovski, ``Nonlinear model reduction in
  power systems by balancing of empirical controllability and observability
  covariances,'' \emph{IEEE Trans. Power Syst.}, vol.~32, no.~1, pp. 114--126,
  Jan. 2017.

\bibitem{vidyasagar2002nonlinear}
M.~Vidyasagar, \emph{Nonlinear Systems Analysis}.\hskip 1em plus 0.5em minus
  0.4em\relax SIAM, 2002.

\bibitem{chow1992toolbox}
J.~H. Chow and K.~W. Cheung, ``A toolbox for power system dynamics and control
  engineering education and research,'' \emph{IEEE Trans. Power Syst.}, vol.~7,
  no.~4, pp. 1559--1564, Nov. 1992.

\bibitem{julier2000new}
S.~Julier, J.~Uhlmann, and H.~F. Durrant-Whyte, ``A new method for the
  nonlinear transformation of means and covariances in filters and
  estimators,'' \emph{IEEE Trans. Autom. Control}, vol.~45, no.~3, pp.
  477--482, Mar. 2000.

\bibitem{hartikainen2011optimal}
J.~Hartikainen, A.~Solin, and S.~S{\"a}rkk{\"a}, ``Optimal filtering with
  kalman filters and smoothers,'' \emph{Department of Biomedica Engineering and
  Computational Sciences, Aalto University School of Science, 16th August},
  Aug. 2011.

\bibitem{CVX1}
M.~Grant and S.~Boyd, ``{CVX}: Matlab software for disciplined convex
  programming,'' Tech. Rep., Sept. 2013.

\end{thebibliography}

\end{document}